\documentclass[pre,preprintnumbers,amsmath,amssymb, twocolumn]{revtex4-1}
\usepackage{graphicx,color}
\usepackage{bm}
\usepackage{epsfig}

\newcommand{\mathds}[1]{{\rm #1 \hspace*{-2mm}#1}}

\graphicspath{{figures_satya/}{figures/}{GNUplot/}{data/}}

\begin{document}

\title{Work Distribution for Unzipping Processes}
\date{\today}

\author{Peter Werner}
\author{Alexander K. Hartmann}	
\affiliation{Institut f\"ur Physik, Universit\"at Oldenburg, 26111 Oldenburg,
  Germany}

\author{Satya N. Majumdar}
\affiliation{Laboratoire de Physique Th\'eorique et Mod\`eles Statistique,
  Universit\'e Paris-Saclay, France}

\begin{abstract}
A simple zipper model is introduced, representing in a simplified way, 
e.g., the folded DNA double helix or hairpin structures in RNA.
The double stranded hairpin is connected to a heat bath at temperature $T$ and subject to an 
external force $f$, which couples to the
free length $L$ of the unzipped sequence. Increasing the force,
leads to an zipping/unzipping first-order phase transition at a critical force $f_c(T)$ in
the thermodynamic limit of a very large chain.
We compute analytically, as a function of temperature $T$ and force $f$, the full distribution
$P(L)$ of free lengths in the thermodynamic limit and show that it is qualitatively very different for $f<f_c$,
$f=f_c$ and $f>f_c$. Next we consider quasistatic
work processes where the force is incremented according to
a linear protocol. Having obtained $P(L)$ already
allows us to derive an analytical expression
for the work distribution $P(W)$ in the zipped phase $f<f_c$ for a long chain. We compute
the large deviation tails of the work distribution explicitly.
Our analytical result for the work distribution
is compared over a large range of the support down to probabilities 
as small as  $10^{-200}$  with
numerical simulations, which were performed by
applying sophisticated large-deviation algorithms.
 \end{abstract}

\maketitle

\section{Introduction}
\label{Intro}

The physical work $W$ plays an important role for
many equilibrium and non-equilibrium processes at all scales. 
The work $W$ is a random variable that fluctuates
from one realization of the underlying process to another.
For large thermodynamic systems, typically the distribution
of work converges to a delta function peaked at its average
and it is thus sufficient to compute just the average work.
For smaller systems, however, the fluctuations of $W$ around
its average are also highly relevant, as illustrated 
by the field of stochastic thermodynamics~
\cite{seifert2012,peliti2021}. Therefore, to describe systems with few degrees of freedom
comprehensively,
one needs to know the full distribution $P(W)$ of the work, and not
just its first moment. For example,
the knowledge of $P(W)$ in a nonequlibrium process connecting two
equilibrium states allows one
to extract the free-energy difference between these two states
by using
the Jarzynski equality \cite{jarzynski1997} or the theorems of Crooks \cite{crooks1998}.

In experiments \cite{collin2005,ciliberto2017}, 
the processes are repeated many times, the work
is measured for each execution, and a histogram of $W$ is obtained.
This allows one to approximate $P(W)$ in the 
high-probability region, i.e., for typical values. But one does not
have access to a broader
range of the support of $P(W)$ down to its tails.
If the system contains few degrees of freedom,
the estimation of free-energy differences without knowing the tails
of $P(W)$ yields wrong results,
because the tails dominate the computed averages of exponentials of work. 
The same is true, when numerical simulations of model processes are performed
in a straightforward way and repeated many times in the so called 
\emph{simple sampling} scheme. Still,
by using \emph{large-deviation algorithms}, the work distribution $P(W)$
has been obtained in few cases 
for a larger range of its support, down to extreme tails with probabilities
as small as $10^{-100}$. In the most interesting case of
complex interacting many-particle systems, $P(W)$ was obtained
for the Ising model subject to a changing
magnetic field \cite{work_ising2014} and for stretching of an 
RNA hairpin structure \cite{Werner_2021}.

By using analytical studies, the distributions $P(W)$ over their
full range of support have
been obtained for
some systems with few degrees of freedom. In very simple cases
Gaussian distributions are obtained \cite{speck2004,speck2005}.
Furthermore, a particle in a cylinder with a moving piston \cite{lua2005}
was considered. Other examples are given by two level systems \cite{quan2008} 
and single 
 particles with a dynamics
described by a Fokker-Planck equation \cite{speck2011,saha2014} or, 
equivalently, 
by a Langevin-equation 
\cite{engel2009,nickelsen2011,nickelsen2012,hoppenau2013,kwon2013,ryabov2013,holubec2015,saha2015,chvosta2020} 
in one-dimensional potentials of varying shapes.
Also single-particle
systems with stochastic driving were 
investigated \cite{verley2014,manikandan2017}. Recently, such approaches
were extended to single-particle quantum systems \cite{funo2018}.
In a related context, models for single-particle heat engines have also been studied 
\cite{hoppenau2013engine,holubec2014,proesmans2015,holubec2022}.

Here we want to go beyond the single-particle case and  
address the analytical calculation of the work distribution $P(W)$
for a more complex system of interacting particles, but with
a finite number $N$ of degrees of freedom. Some works
have been done in that direction.
For example, in a system of several point particles coupled by harmonic springs, e.g..,
the Rouse polymer model, the
work distribution can be obtained from the result of the
single-particle Gaussian case \cite{speck2005}. 
In addition to the harmonic coupling also the extreme case of very
stiff nonlinear coupling can be solved \cite{vucelja2015}.
The treatment of polymers can be extended to networks of harmonic oscillators
\cite{damak2020}.
Furthermore, for a relaxing elastic
manifold \cite{wu2023} and for a chain of active particles coupled by
harmonic springs
\cite{gupta2021} the heat statistics have been obtained analytically.

Our present work is motivated
by the recent numerical study on stretching of RNA
secondary structures, where $P(W)$ was obtained for hairpins.
By applying large-deviation algorithms, 
almost the full range of the support could be obtained \cite{Werner_2021}.
In the present study we will investigate a simplified \emph{zipper} model
that can be used to describe either
the opening and closing of RNA hairpin structures, or that of
DNA double helices,  under an external 
force. Our zipper model is 
a very simple one with non-harmonic interactions
(and thus goes beyond Gaussian integrals in models
with harmonic interactions), and yet 
allows for an exact analytic solution in the folded or `zipped' phase, in particular
exhibiting a non-Gaussian distribution of work $P(W)$.

In an early model for DNA unzipping in a solvent \cite{gibbs1959},
the term coupling the
free part of the helix to a solvent can be interpreted as the interaction
of the free part with an external force $f$.  
The model allowed for an exact calculation of the partition
function and an unzipping transition was observed.
During the decades other variants of
unzipping 
models \cite{applequist1963,bockelmann1997,marenduzzo2001} were studied,
like the unzipping of DNA by pulling \cite{lubensky2000}, 
and the denaturation of DNA under applied torque \cite{hwa2003}.
Also a model exhibiting a first-order unzipping
transition was analyzed \cite{Kafri2000,Kafri2002,Kafri2002b}.
Furthermore, unzipping was studied via the calculation of Lee-Yang
zeros \cite{Deger2018}, which allowed the authors to obtain the
large-deviation tail of the energy distribution.
In some cases heterogeneous sequences were considered \cite{roland2009},
in particular a mapping to the disordered polymer in random media
was provided \cite{Kafri2006}, allowing for an analytical replica
calculation.
The actual dynamics of unzipping was modeled via a simple kinetic
two-step process \cite{sauer-budge2003},
for reviews see, e.g., Refs.~\cite{cocco2002,rissone2022}. 
For the equilibrium  models, 
unzipping transitions with model-dependent 
critical forces $f_{\rm c}$ were found
and the behaviors in the zipped and the unzipped phases were described
by average quantities (and sometimes also by the corresponding
fluctuations). One important observable
is the free length $L$, i.e., the unzipped
part of the sequence which is next to the beginning of the sequence
where the external force is applied.

In the present paper, we will use a simple version of the previous
models but go beyond the calculation of 
averages and study the full distribution $P(L)$ of the free length
in the zipped phase, in the unzipped phase, and also at the critical force 
$f_{\rm c}$.
This in turn will allow us to calculate analytically 
some moments and finally the full distribution $P(W)$ of work for equilibrium
 processes where the external force is changed from $0$ to $f_{\max}$ quasistatically,
in the region $f<f_{\rm c}$ (zipped phase).  We
compare the analytical results to large-deviation simulations of the unzipping process
and find very good agreement over about 200 decades in probability.
 Thus, the model constitutes a nice example
where the work distribution is available for a process in a system
of many interacting particles. This can be used as a starting
point for similar consideration of the corresponding
not-quasistatic process or other models of complex interacting particles.
Also the shape or generalizations of the obtained 
distribution can be used to fit to numerical data for other systems.

The paper is organized as follows. In Sec.~\ref{Model}, we
present our zipper model. Then in Sec.~\ref{Equilibrium}, we solve the model exactly and show
the existence of a first-order unzipping transition in the
temperature-force plane and obtain
in particular the full distribution $P(L)$ of the free length. In 
Sec.~\ref{sec:sampling} we present an exact numerical algorithm to sample
configurations in equilibrium. In Sec.~\ref{numeric:transition}, we
use this algorithm to confirm the analytical results for the
thermodynamic behavior. Next, in Sec.~\ref{analytic:work_distr},
we analytically derive, in the zipped phase, 
the first two moments and also the full distribution of work $P(W)$ for
quasistatic processes involving a finite number of increments of the force $f$.
In Sec.~\ref{sec:largeN_s}, we evaluate these quantities in the limit of infinitely large
increments. Then, in Sec.~\ref{num:large_deviation}, we present the
numerical results of the large-deviation 
simulations showing the work distributions 
over hundreds of decades in probability and compare with the analytical
findings. Finally, we conclude in Sec.~\ref{Conclusion} with a summary and outlook.

\section{A simple unzipping model\label{Model}}

\begin{figure}
\begin{center}
\includegraphics[width=0.7\columnwidth]{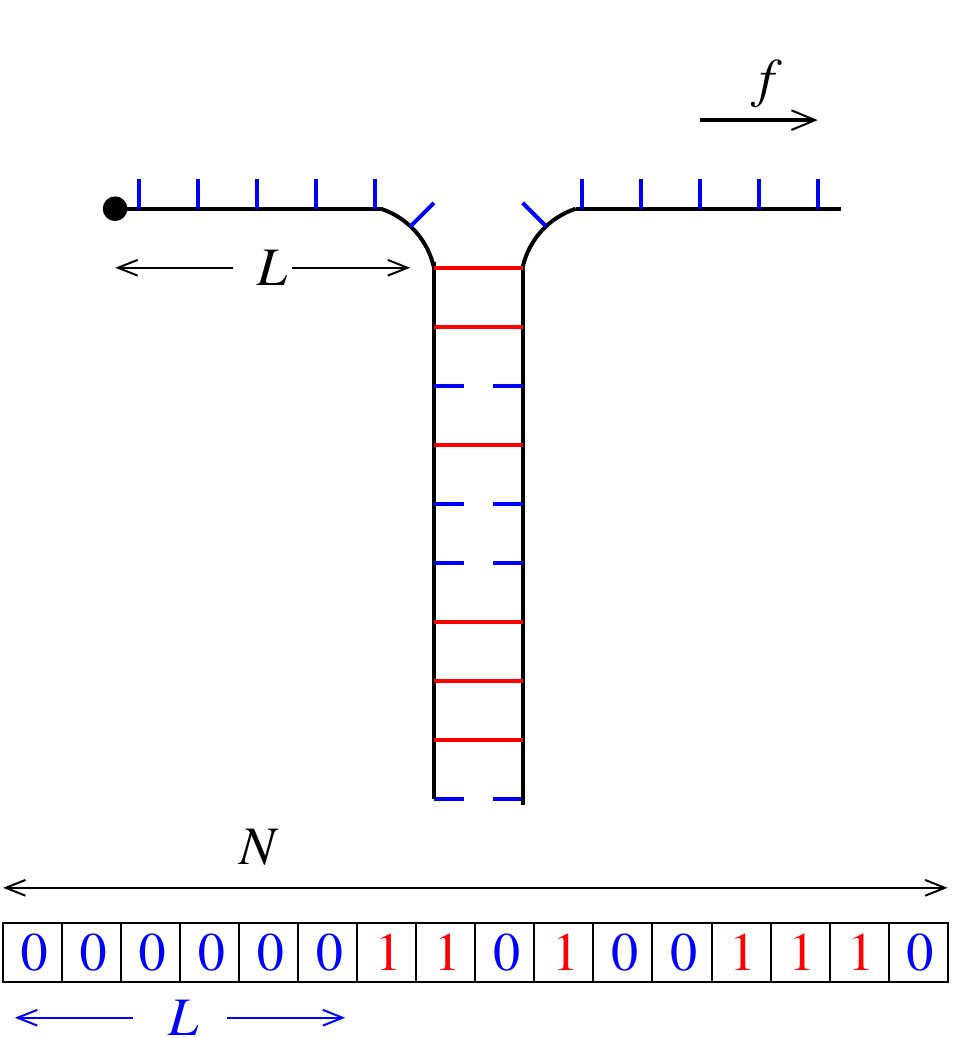}
\end{center}
\caption{Top: A sequence which is partially unzipped under an applied
  force $f$. The length of the outside unzipped part is $L$.
Also in the zipped part, some pairs are not bonded, forming so called
bubbles.
Bottom: binary-string representation of length $N$ of this configuration
consisting of  $1$'s and $0$'s. Thus,  $L$ denotes
the length of the substring made of consecutive $0$'s from the left
before the first $1$.
\label{fig1.rna}}
\end{figure}

We consider a zipper consisting of two complementary
halfs of a sequence consisting of $N$ opposite pairs of bases, where each pair can
independently be bonded (close) or unbonded, see the top part of Fig.~\ref{fig1.rna}.
Note that we assume the simplest case
where each base can only be bonded to the complementary base
in the pair, not to other bases.

The part of the sequence
that is outside the ``outmost'' bonded pair, i.e., the upmost pair in 
the top part of Fig.~\ref{fig1.rna},  is denoted as \emph{free}.
We consider the
 case where the first base is fixed and the last base is coupled to, e.g., an
optical tweezer, such that a force $f$ 
can be exerted on the zipper. Thus, it is the
free part of the zipper which couples to the force.

This situation can be described by a
one dimensional lattice of $N$ sites $i=1,\ldots,N$,
which represents the first 
half or strand of the zipper. At each site, we have a binary variable
$\sigma_i\in \{0,1\}$, indicating whether the base $i$ is bonded
($\sigma_i=1$) or not,
respectively. A typical configuration of this binary string is shown in 
Fig.~\ref{fig1.rna}. Let $M$ denote the total number of $1$'s in a configuration.
We also denote by $L$ the number of lattice sites to the left of the first $1$ (appearing
in the string as one reads from left to right)--this is one half of 
the free length of the zipper.
Evidently all these $L$ sites contain $0$
by definition, see bottom of Fig.~\ref{fig1.rna}. 
Thus a typical configuration is labeled by two integers
$M$ and $L$. Note that for a given $L$ in $0\le L\le N$, the number of $1$'s, i.e., the
variable $M$ can take values only in the range
$0\le M\le N-L$. We define the energy of the configuration as
\cite{mueller2002}
\begin{equation}
E(M,L)= - J\, M - 2\, f\, L
\label{energy.1}
\end{equation}
where $f>0$ is the applied force and $J>0$ is the binding energy
of a base pair between a base in one strand
and its partner base located at the equivalent position in the other strand.

\section{Equilibrium behavior}
\label{Equilibrium}

\noindent {\bf Ground state:} Let us first investigate the ground state by minimizing
the energy function in Eq.~(\ref{energy.1}). We need to maximize $-E(M,L)=J\, M+ 2\, f\, L$.
Since both terms in $-E$ are non-negative, we can maximize them one after
the other. First fix $L$
and vary $M$. The maximum value of $M$, for fixed $L$, is clearly $(N-L)$. Hence
$-E(M=N-L,L)= J\,(N-L)+2\,f\, L= J\,N+ (2f-J)\, L$. We now have to maximize 
this function with respect to $L$ where $0\le L\le N$. There are two possibilities.

\begin{itemize}

\item The case $0<f<J/2$: In this case, the function $J\,N+ (2f-J)\, L$ is maximized when $L=0$,
i.e., the first entry (from the left) in the ground state configuration must be a $1$. This
is then the `zipped' phase.

\item The case $f>J/2$: In this case, the function $J\,N+ (2f-J)\, L$ is maximized when $L=N$.
This means the ground state consists of all $0$'s. This is thus a totally `unzipped' phase.

\end{itemize} 

In summary, there is a phase transition in the ground state from the `zipped' phase
to the `unzipped' phase, as one increases $f$ (fixed $J$)
through the critical value $f_c= J/2$. We will see below that this `unzipping' phase transition
persists at finite temperature also.

\vskip 0.5cm

\noindent {\bf Finite temperature:} At finite temperature, we associate a Boltzmann weight
$e^{-\beta\, E(M,L)}$ to each configuration labeled by $(M,L)$ where the energy $E(M,L)$
is given in Eq.~(\ref{energy.1}) and $\beta= 1/T$ is the inverse temperature. Henceforth,
we fix $J$ and consider the behavior of the system as a function of two control
parameters $(T, f)$ in the force-temperature plane. Our goal is to obtain the
phase diagram in the $(T-f)$ plane. The
partition function of the model is defined as
\begin{equation}
Z_N(\beta,f)= \sum_{\rm all\,\,\, config.} e^{-\beta\, E(M,L)}\, .
\label{pf.1}
\end{equation}
To evaluate the partition function, we carry out the sum over configurations
in two steps. We first fix $L$ and sum over all values of $M$. After this,
we sum over all values of $L$. Thus we write
\begin{equation}
Z_N(\beta,f)= \sum_{L=0}^N W_N(L)
\label{pf_step1}
\end{equation}
where we define
\begin{equation}
W_N(L)= \sum_{M=0}^{N-L} e^{-\beta\, E(M,L)}\, .
\label{def_WNL}
\end{equation}

When carrying out the sum over $M$, we need to distinguish two cases, namely
when $0\le L\le N-1$ and when $L=N$. In the latter case ($L=N$), the string 
consists entirely of $0$'s, hence $M=0$. In this case, the Boltzmann weight factor
is just $e^{2\beta f N}$. For $0\le L\le N-1$, we note that once we fix $L$, the
$(L+1)$-th entry is necessarily a $1$. Hence, the remaining $(M-1)$ $1$'s can
be placed in the available $(N-L-1)$ boxes such that each box can contain at most
one $1$. The number of ways this can be done is simply
${N-L-1 \choose M-1}$. Hence the net partition sum obtained by summing over all
possible $M$, for fixed $0\le L\le N$, can be expressed as
\begin{multline}
W_N(L)= e^{2\beta f N}\, \delta_{L,N} + \\
\left[\sum_{M=1}^{N-L} {N-L-1 \choose M-1} \, e^{\beta J M+ 2\beta f L}\right]\, 
{\mathds{1}}_{0\le L\le N-1} 
\label{WL.1}
\end{multline}
where $\mathds{1}_{0\le L\le N-1}$ is an indicator function which is one
if $0\le L \le N-1$ and $\delta_{L,N}$ is the Kronecker delta function.
The sum over $M$ in Eq.~(\ref{WL.1}) can be performed trivially using binomial expansion and 
one gets, after slight
rearrangement,
\begin{multline}
W_N(L)= \\
\frac{(1+e^{\beta J})^N}{(1+e^{-\beta J})}\,\left[ \mu^L\,  
{\mathds{1}}_{0\le L\le N-1}  + (1+e^{-\beta J})\, \mu^N\, \delta_{L,N}\right]\, ,
\label{WL.2}
\end{multline}
where we introduced the important parameter
\begin{equation}
\mu= \frac{e^{2\beta f}}{1+e^{\beta J}}\, .
\label{mu_def}
\end{equation}
Finally, summing Eq.~(\ref{WL.2}) over $L$ (it is just a simple geometric series), 
we get the exact partition function
valid for arbitrary positive $N$
\begin{multline}
Z_N(\beta,f)= \sum_{L=0}^N W_N(L)= \\
\frac{(1+e^{\beta J})^N}{(1+e^{-\beta J})}\,\left[
\frac{1-\mu^N}{1-\mu} + (1+e^{-\beta J})\, \mu^N\right]\,  
\label{pf}
\end{multline}
where $\mu$ is given in Eq.~(\ref{mu_def}).

We now analyze the thermodynamic limit $N\to \infty$. The free energy per site is defined
as
\begin{equation}
{\cal F}(\beta,f)= -\lim_{N\to \infty}\frac 1 \beta \frac{\ln Z_N(\beta,f)}{N}\, .
\label{fe.1}
\end{equation}
Taking logarithm of Eq.~(\ref{pf}) and the $N\to \infty$ limit, 
we see that the limiting value of the free energy per site depends
on whether $\mu\le 1$ or $\mu>1$. We obtain
\begin{eqnarray}
\label{fe.2}
{\cal F}(\beta,f)=\begin{cases}
&-\frac 1 \beta \ln (1+e^{\beta J}) \quad {\rm if}\quad \mu\le 1  \\
\\
&- 2\, f \quad\quad\quad\quad\, {\rm if}\quad \mu>1 \, .
\end{cases}
\end{eqnarray}
Note that $\cal F$ does not depend on the force for $\mu<1$ and
is continuous at $\mu=1$, but its first derivative with respect to
$T$ or $f$ is discontinuous at the critical point $\mu=1$, indicating a first-order
phase transition. 

\begin{figure}
\includegraphics[width=0.99\columnwidth]{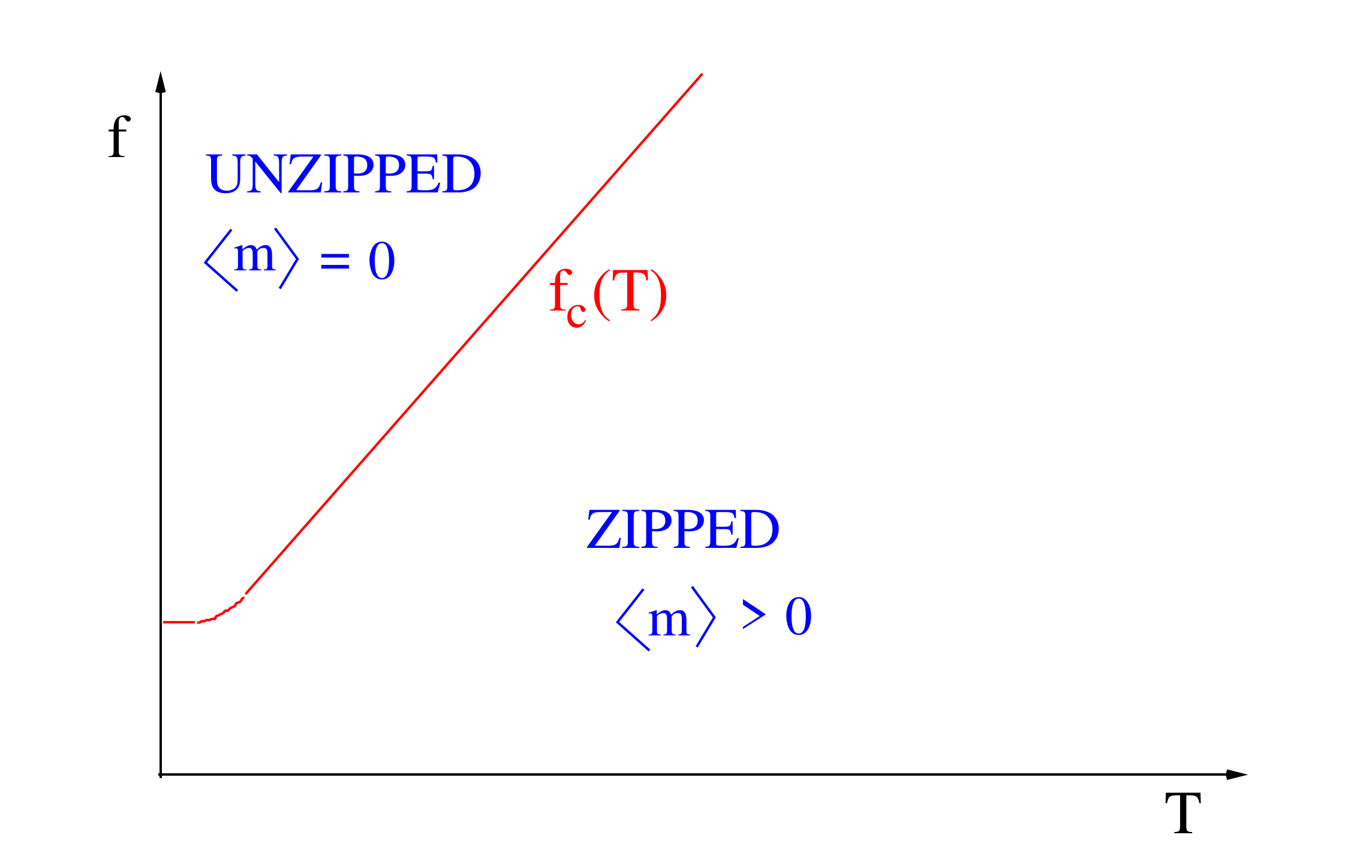}

\caption{Phase diagram in the $(T-f)$ (temperature-force) plane.
The critical line $f_c(T)= (T/2)\, \ln(1+e^{J/T})$  (drawn schematically in the figure) 
separates the
unzipped phase ($f>f_c(T)$) from the zipped phase ($f<f_c(T)$).
The order parameter $\langle m\rangle$, namely the fraction of $1$'s in
the string in the thermodynamic limit, is nonzero in the zipped phase and
vanishes in the unzipped phase. On the critical line $f=f_c(T)$, the
order parameter $\langle m\rangle$ is nonzero and jumps to $0$ as
one enters the unzipped phase from the zipped side, indicating a
first-order phase transition.
\label{fig.phd}}
\end{figure}

To shed more lights on the two phases and the transition between them,
let us now define the average fraction of $1$'s in the string as an order parameter
\begin{equation}
\langle m\rangle= \lim_{N\to \infty} \frac{\langle M\rangle}{N}= - 
\frac{\partial {\cal F}(\beta,f)}{\partial J}\, .
\label{op.1}
\end{equation}
Using Eq.~(\ref{fe.2}) for the free energy per site, one gets
\begin{eqnarray}
\label{op.2}
\langle m\rangle=\begin{cases}
& \frac{e^{\beta J}}{1+e^{\beta J}}  \quad {\rm if}\quad \mu\le 1  \\
\\
& 0 \quad\quad\quad\, {\rm if}\quad \mu>1 \, .
\end{cases}
\end{eqnarray}
Thus the phase $\mu>1$ corresponds to the `unzipped' phase where $\langle m\rangle=0$,
while the phase $\mu\le 1$ corresponds to the `zipped' phase where $\langle m\rangle$
is nonzero. As one approaches the critical point $\mu\to 1$ from below, the
order parameter $\langle m\rangle$ undergoes a finite jump, thus confirming
the first-order phase transition,

Thus the critical line in the $(T-f)$ plane is obtained by setting $\mu=1$. Using
the expression of $\mu$ in Eq.~(\ref{pf}) one then obtains the critical curve
$f_c(T)$ in the $(T-f)$ plane
\begin{equation}
\mu=1 \Longrightarrow f_c(T)= \frac{\ln (1+e^{\beta J})}{2\beta}= \frac{T}{2}\, \ln(1+e^{J/T})\, .
\label{fc_T.1}
\end{equation}
The phase diagram in the $(T-f)$ plane including the critical line $f_c(T)$ is shown in
Fig.~\ref{fig.phd}. Note that the critical force \emph{increases}
when increasing the temperature, i.e., the fluctuations do not
help. This is known as \emph{cold unzipping} in the literature.

The critical curve $f_c(T)$ has the following asymptotic behaviors
\begin{eqnarray}
\label{crit_line}
f_c(T)\simeq \begin{cases}
& \frac{J}{2}+ \frac{T}{2}\, e^{-J/T}- \frac{T}{4}\, e^{-2J/T}+\cdots  
\quad (T\to 0) \\
\\
& \frac{1}{2}(\ln 2)\, T+ \frac{J}{4}+ \frac{J^2}{16T} +O(T^{-2}) \quad 
(T\to \infty) \, .
\end{cases}
\end{eqnarray}
To obtain the $T\to 0$ limit, we rewrite the expression of $f_c(T)$ in
Eq. (\ref{fc_T.1}) as $f_c(T)= J/2+ (T/2) \ln (1+ e^{-J/T})$ and
then expand the logarithm in powers of $e^{-J/T}$. This gives
the first line of Eq. (\ref{crit_line}). In the opposite $T\to \infty$ limit, we
first expand $e^{J/T}= 1+J/T+ J^2/(2T^2)+\ldots$ and then expand the logarithm
in Eq. (\ref{fc_T.1}) in powers of $1/T$, yielding the second line
in Eq. (\ref{crit_line}).
Thus, as $T\to 0$, the critical value $f_c(0)=J/2$ is consistent with the ground state
analysis before.

\vskip 0.5cm

\noindent{\bf Statistics of the free length $L$:} One can also characterize the
transition in terms of the variable $L$, denoting the free
length of the RNA chain.
Here, we will not only calculate the average value but actually
the full distribution
$P(L|N)$ for any given $N$. This will allow us later on to obtain
for slow processes the full distribution of work analytically.

\begin{figure}[ht]
  \begin{center}
    \includegraphics[width=0.8\columnwidth]{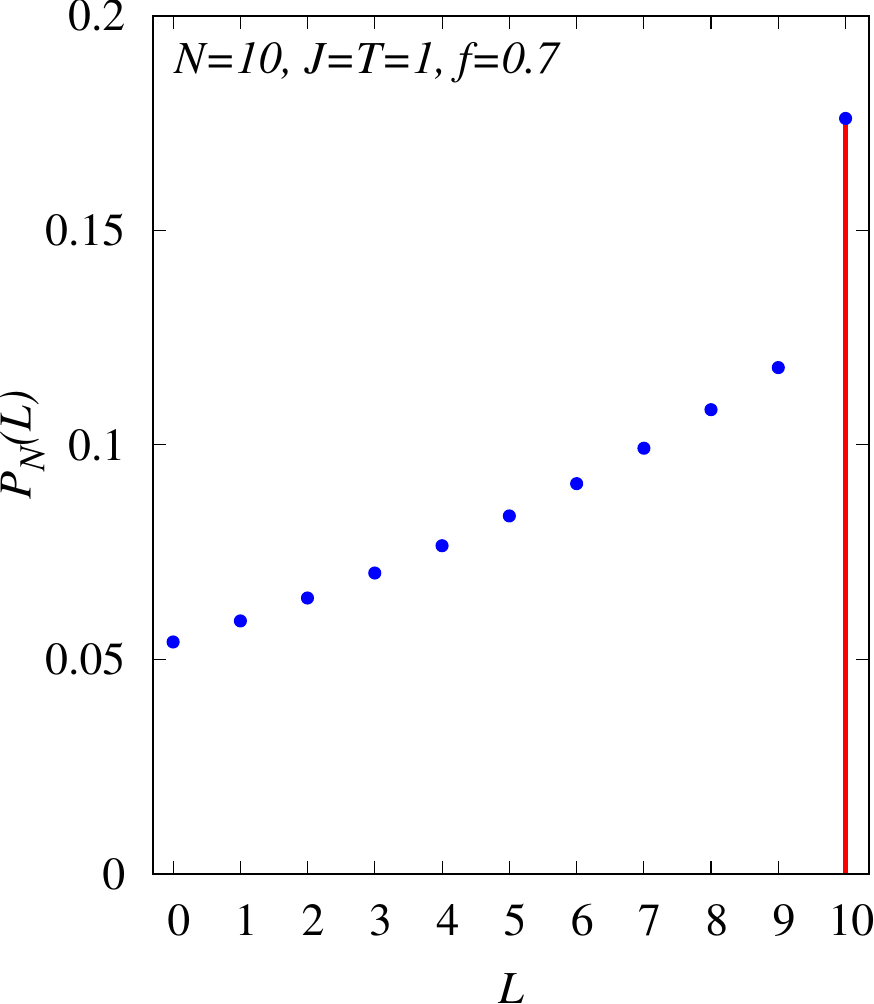}
    \end{center}
\caption{\label{fig:distr:unzipped:phase}
  Distribution of the free length $L$ in the
  unzipped phase (see Eqs.~(\ref{pnl.2})) for
  $T=J=1$, $N=10$ and force value $f=0.7>f_c$, where $f_c\approx0.6566$.
  The red vertical line highlights the delta peak at $L=N$.}
\end{figure}

Indeed, it follows from Eq.~(\ref{WL.2}) that $P(L|N)$ is 
given exactly, for all $0\le L\le N$, by
\begin{multline}
P_N(L)= \frac{W_N(L)}{Z_N(\beta,f)}= \\
\frac{ \mu^L\, {\mathds{1}}_{0\le L\le N-1} +
(1+e^{-\beta J})\, \mu^N\, \delta_{L,N} }{\frac{1-\mu^N}{1-\mu}+ (1+e^{-\beta J})\, \mu^N} \, .
\label{pnl.2}
\end{multline}
It is easy to check that $P_N(L)$ is normalized, i.e., $\sum_{L=0}^N P_N(L)=1$.
A plot of this distribution for a small value of $N$ and $f>f_c$
  highlighting the delta peak is given in Fig.~\ref{fig:distr:unzipped:phase}.

From Eq.~(\ref{pnl.2}), it is easy to compute its first moment for all $N$
\begin{multline}
\langle L\rangle= \sum_{L=0}^N L\, P_N(L)=\\ 
\frac{ \frac{\mu}{1-\mu}\left[\frac{1-\mu^N}{1-\mu}-N\, \mu^{N-1}\right]
+ (1+e^{-\beta J})\, N\, \mu^N}{\frac{1-\mu^N}{1-\mu}+ (1+e^{-\beta J})\, \mu^N}\, ,
\label{avgl.1}
\end{multline}
where $\mu$ is given in Eq.~(\ref{mu_def}). In Sec.~\ref{numeric:transition}
samples for $\langle L \rangle$ as function of the force $f$
are shown and compared to result from numerical exact sampling.

It is interesting to compute $\langle L\rangle$ in the large $N$ limit. It follows
from Eq.~(\ref{avgl.1}) that this thermodynamic limit depends crucially on
whether $\mu>1$, $\mu<1$ or $\mu=1$. Taking this limit carefully, we find that
as $N\to \infty$
\begin{eqnarray}  
\label{avgl.2}
\langle L\rangle \simeq \begin{cases}
& \frac{\mu}{1-\mu} \quad\quad {\rm for}\quad\quad \mu<1  \\ 
\\
& \frac{N}{2} \quad\quad\quad {\rm for}\quad\quad \mu=1  \\
\\
& N \quad\quad\quad {\rm for}\quad\quad \mu>1 \, .
\end{cases}
\end{eqnarray} 
Hence, as $N\to \infty$, the average free length available per site $\langle l\rangle= \langle L\rangle/N$
approaches to $0$ for $\mu<1$ (zipped phase) and $1$ for $\mu>1$ (unzipped phase).
Exactly, on the critical line $\mu=1$, we find $\langle l\rangle\to 1/2$. This
statistics of $\langle l\rangle$ also confirms the first-order
nature phase transition at $\mu=1$ for a different measurable quantity.

Let us also analyze the asymptotic form of the full distribution $P_N(L)$ in Eq.~(\ref{pnl.2})
in the
thermodynamic limit $N\to \infty$. The behavior again depends on whether $\mu< 1$ (zipped phase),
$\mu>1$ (unzipped phase) or on the critical lime $\mu=1$ (critical).

\begin{itemize}

\item Zipped phase ($\mu< 1$): In this case, as $N\to \infty$ in Eq.~(\ref{pnl.2})
  the distribution $P_N(L)$ converges to an $N$ independent form which is purely geometric.
  For $L=0,1,2,\cdots$ one obtains
\begin{equation}
P_{N\to \infty}(L)= (1-\mu)\, \mu^L \, .
\label{pl_zipped.1}
\end{equation}
In this $N\to \infty$ limit, the average value $\langle L\rangle $ approaches a constant
\begin{equation}
\langle L\rangle= (1-\mu)\sum_{L=0}^{\infty} L\, \mu^L= \frac{\mu}{1-\mu}\, ,
\label{Lavg.1}
\end{equation}
in accordance with the first line of Eq.~(\ref{avgl.2}).
The fluctuations of $L$
around this mean is quantified by the variance ${\rm Var}(L)= \langle L^2\rangle- \langle L\rangle^2$
that also approaches to a constant ${\rm Var}(L)= \mu/(1-\mu)^2$ in the large $N$ limit.

\item Critical line ($\mu=1$): On the critical line $\mu=1$, we find that
  $P_N(L)$ in Eq.~(\ref{pnl.2}) approaches a scaling form as $N\to \infty$
\begin{equation}
P_{N\to \infty}(L) \to \frac{1}{N}\, F\left(\frac{L}{N}\right)\, ,
\label{pl_crit.1} 
\end{equation}
where the scaling function $F(x)=1$ for $0\le x\le 1$ and $F(x)=0$ for $x>1$. In other words,
the distribution of $L$ approaches a flat uniform distribution over $L\in [0,N]$. 
Consequently, the average value $\langle L\rangle$ approaches the value $N/2$ as found
in the second line of Eq.~(\ref{avgl.2}). One can also check that the variance
${\rm Var}(L)= \langle L^2\rangle- \langle L\rangle^2 \to N^2/12$ as $N\to \infty$, indicating
that the fluctuations of $L$ remain big in the large $N$ limit at the critical point, unlike
in the zipped phase where it is of $O(1)$ as $N\to \infty$.

\item Unzipped phase ($\mu>1$): In this phase, the distribution $P_N(L)$ in Eq.~(\ref{pnl.2}) 
does not approach
a limiting form as $N\to \infty$. Instead, there is a decaying $N$-dependent `bulk' part that co-exists
with a delta peak (condensate) at the right edge of the support at $L=N$. More precisely, for large $N$, 
we find from Eq.~(\ref{pnl.2}) that
\begin{eqnarray} & &
P_{N\to \infty}(L)  \to \frac{(\mu-1)}{\left[1+(\mu-1) (1+e^{-\beta J})\right]}
\times \nonumber \\ 
&& \left[ \mu^{L-N}\, {\mathds{1}}_{0\le L\le N-1} + (1+e^{-\beta J}) \, \delta_{L,N}\right]\, .
\label{pl_unzipped.1}
\end{eqnarray}
Note that the weight of the delta peak or the condensate at $L=N$ approaches an $N$-independent value as
  $N\to \infty$.
However, the `bulk' part of the distribution, even though decays
exponentially as $\sim \mu^{L-N}$ away from the condensate, does actually contribute to all moments of $L$ even in
the $N\to \infty$ limit. For example, the average value $\langle L\rangle \to N + O(1)$ (as $N\to \infty$)
as in the third line of Eq. (\ref{avgl.2}). The variance ${\rm Var}(L)= \langle L^2\rangle-\langle L\rangle^2$
however approaches to a constant as $N\to \infty$
\begin{eqnarray}
{\rm Var}(L)\to \frac{\mu \left[1+ (\mu^2-1)(1+e^{-\beta J})\right]}{(\mu-1)^2 \left[1+ (\mu-1)(1+e^{-\beta J})\right]^2}\, .
\label{varL.unzipped}
\end{eqnarray}  
For both moments, the delta peak as well as the `bulk' part contributes in the large $N$ limit.
So, one can not neglect this `bulk' part in the thermodynamic limit.
This is thus a rather unusual and interesting distribution.

\end{itemize}

The limiting distribution is compared for different values of $f$
to the ones obtained from
numerical exact sampling in Sec.~\ref{numeric:transition}. The numerical
approach is described next.

\section{Sampling of configurations\label{sec:sampling}}

To sample configurations numerically, we calculate for a given system
length $N$ two conditioned partition functions $Z_f(i)$ and $Z_z(i)$, $i=1,\ldots,N$.
Here
$Z_f(i)$ is the partition function of the subsequence $i \ldots N$
conditioned
to the case
that all sites $1,2\ldots i-1$ are not bonded, i.e., on the free part. $Z_z(i)$
is the corresponding partition function for subsequence $i \ldots N$
for the zipped case, i.e., for at least one site $j\in\{1, \ldots,i-1\}$
we have $\sigma_i=1$. The simplest case is for $i=N$, i.e., the sequence
is only the last pair. If the preceding subsequence
$\sigma_1,\ldots \sigma_{N-1}$  is free, having the pair open will contribute
according to Eq.~(\ref{energy.1}) the energy $-2f$ while closing the
pair will contribute energy $-J$. Correspondingly if the preceding
subsequence is not fully free, there will be no coupling
to the force in the final site $i=N$. Therefore,
having the pair open contributes energy 0,
while closing the pair again yields energy $-J$. Thus,
one obtains
\begin{eqnarray}
Z_f(N) & = & e^{2f/T}+e^{J/T} \nonumber \\ 
Z_z(N) & = & e^{0}+e^{J/T}   \,.
\end{eqnarray}
This can be used as starting point
for the
recursive equations which read for $i\in\{0,\ldots,N-2\}$:
\begin{eqnarray}
Z_f(i) & = & e^{2f/T} Z_f(i+1)+ e^{J/T}Z_z(i+1) \nonumber \\
Z_z(i) & = & (e^0+e^{J/T})Z_z(i+1) = (e^0+e^{J/T})^{N-i} \,.
\end{eqnarray}
Thus, the computation is done by the \emph{dynamic programming} approach
\cite{practical_guide2015},
by starting at $i=N$ and iterating site $i$ until $i=1$
is reached. Thus, the calculation takes $O(N)$ steps.
The case $i=1$ describes the full sequence
and therefore the complete partition function is given by $Z_f(1)$,
while $Z_z(1)$ is not used.
Note that here it is rather easy to introduce site randomness by
generalizing $J \to J_i$. 

To sample a configuration $\sigma_1,\ldots,\sigma_{N}$
one starts at site $i=1$, in the free part
of the chain, and assigns iteratively variables $\sigma_i$,
$i=1,\ldots,N$.
As long as the partial configuration is free,
i.e., $\sigma_j=0$ for all $j<i$,
one assigns $\sigma_i=0$
with probability $p_f^0(i)=e^{2f/T}Z_f(i+1)/Z_f(i)$. Thus,
with probability $1-p_f^0(i)$ one assigns $\sigma_i=1$. Once the first
non-zero assignment $\sigma_i=1$ has been made, one has reached the zipped
part of the chain. From now, on assigns $\sigma_i=0$
with probability $p_z^0(i)=Z_z(i+1)/Z_z(i)=1/(1+e^{J/T})$ and
$\sigma_i=1$ with probability $1-p_z^0(i)=e^{J/T}/(1+e^{J/T})$.
Note that this sampling is performed in perfect equilibrium,
and all sampled configurations are statistically independent,
for arbitrary values of temperature $T$ and force $f$. To sample
one configuration it takes a linear $O(N)$ number of steps. While
sampling, one can directly record the size $L$ of the free length and
the number $M$ of bonded sites.

\section{Numerical results for the unzipping transition \label{numeric:transition}}

\begin{figure}[ht]
\includegraphics[width=0.99\columnwidth]{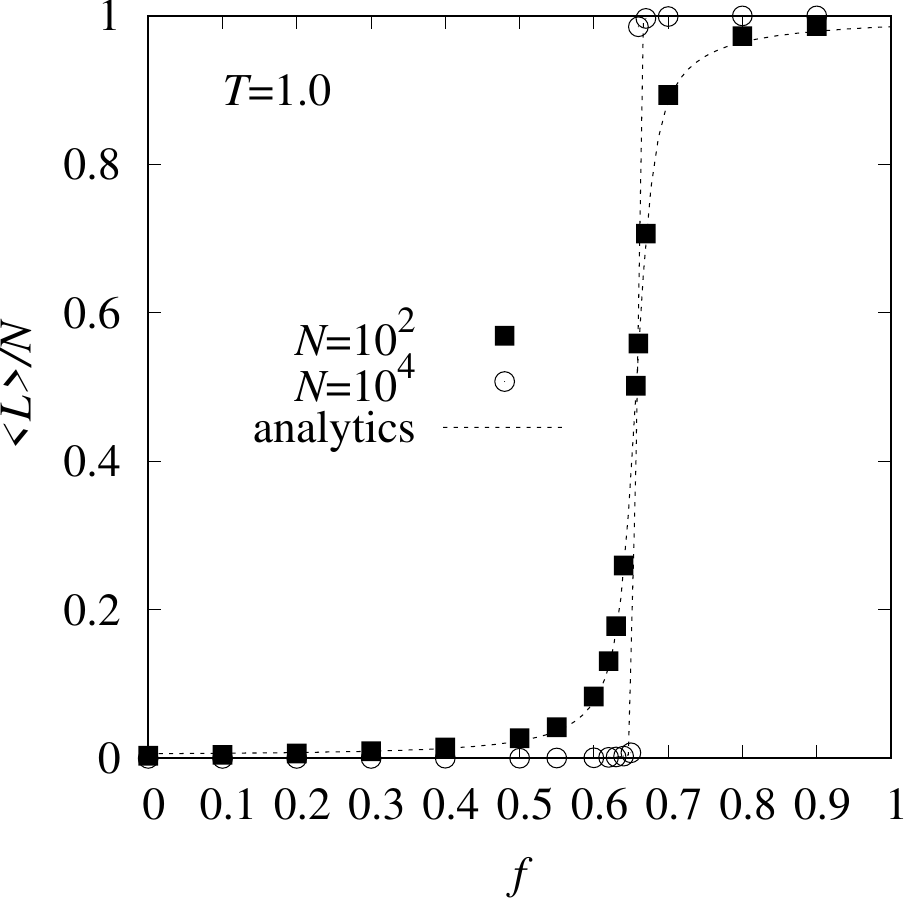}
\caption{\label{fig:L_f_N100}
  Normalized expectation value $\langle L \rangle /N$ of the
  length of the unzipped part as a function of the force strength $f$,
  for temperature $T=J=1$ and two different system sizes $N=10^2,10^4$.
  The symbols denote the numerical results, while the lines the
  analytical results from Eq.~(\ref{avgl.1}). }
\end{figure}

In Fig.~\ref{fig:L_f_N100} the result for the average length
$\langle L \rangle$ is shown for $T=J=1$ as a function of the force
strength $f$. To investigate the finite-size effects, two different
system sizes $N=10^2$ and $N=10^4$ are displayed. For the numerical
result, an average over $10^6$ randomly sampled configurations
was taken for $N=10^2$, while $10^4$
configurations were considered for $N=10^4$. An excellent agreement with the
analytical results from Eq.~(\ref{avgl.1}) is observed. In particular
the first-order nature of the transition is very well visible for $N=10^4$.

\begin{figure}[ht]
\includegraphics[width=0.99\columnwidth]{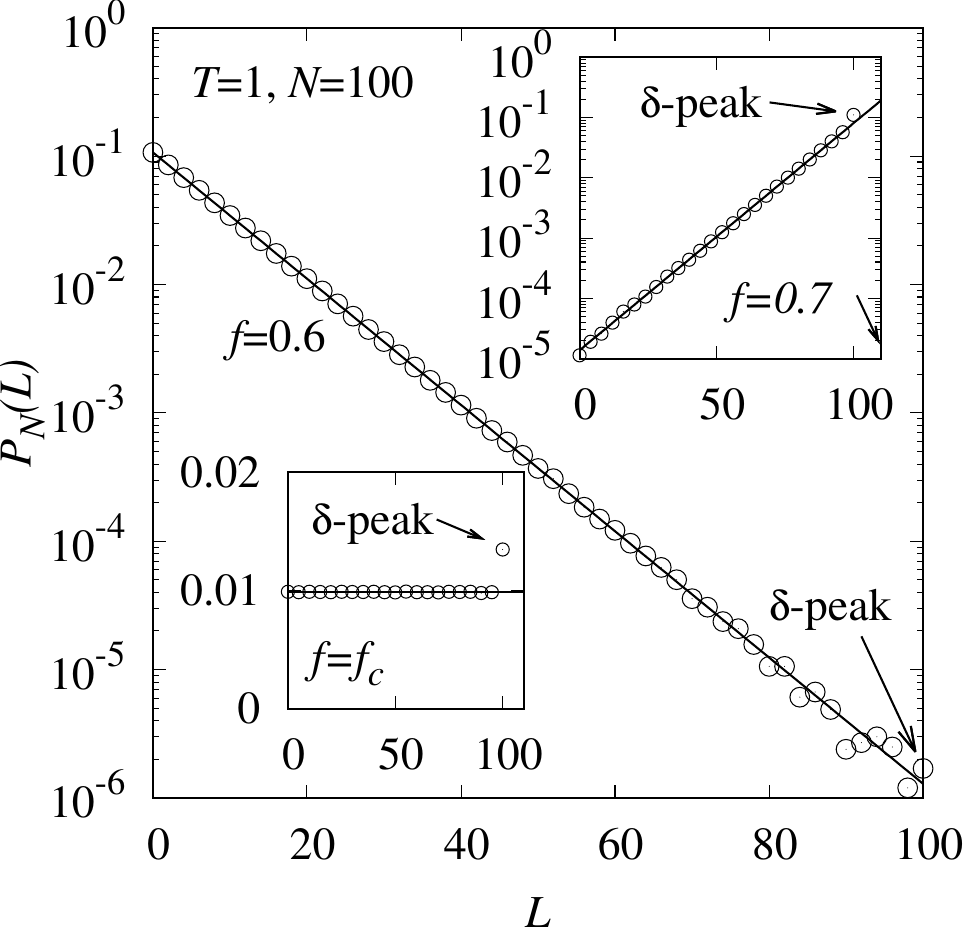}
\caption{\label{fig:distr} Distribution of the free length $L$ for
  $T=J=1$, $N=100$ and
  for force values $f=0.6<f_c$, $f=f_c\approx0.6566$ and $f=0.7>f_c$,
  respectively. The symbols denote numerical results, while the lines
  show the analytical asymptotic distributions
  from Eqs.~(\ref{pl_zipped.1}), (\ref{pl_crit.1}) and (\ref{pl_unzipped.1}).
The delta peak at $L=N$
  is for some cases only slightly visible, in particular for $f=0.6$ where
  the relative flucations are large,
  and therefore highlighted in the figure by arrows,
respectively.}
\end{figure}

In Fig.~\ref{fig:distr} examples for the distributions $P_N(L)$ of the
free length  are shown, for the case $T=J=1$ and $N=100$.
Three cases $f<f_c$, $f=f_c$ and $f>f_c$ are presented.
  The numerical results are histograms obtained from  $10^7$
  independently sampled configurations.
  For $f<f_c$ a clear decreasing
  exponential function is visible. Despite the rather small
  system size $N=100$, a very good match with the limiting $N\to \infty$
  analytical result from Eq.~(\ref{pl_zipped.1}) is visible, apart from
  the statistical fluctuations for large values of $L$. For
$f=f_c$ an almost
  full uniform distribution is found, as obtained in Eq.~(\ref{pl_crit.1}),
  plus a small peak at $L=100$, which is present in the full
  distribution Eq.~(\ref{pnl.2}), but should decrease in weight for
  $N\to\infty$.
  For $f>f_c$
  a rising exponential, matching the result of Eq.~(\ref{pl_unzipped.1})
  is clearly visible. Here also a  
  finite peak at $L=N$ appears, which should remain also for large system
  sizes. In general,
  finite-size effects appear very small.
  Thus, the rather small size $N=100$ almost represents the limiting
  $N\to\infty$ behavior.

\section{Work distribution in the zipper model in the zipped phase
\label{analytic:work_distr}}

In this section, we compute
the work distribution in the zipper model in
the zipped phase by increasing the applied force quasistatically. Here, we will consider 
increasing
the applied force in discrete steps in units of $f_0>0$.
In other
words, the applied force at the $m$-th step is given by
\begin{equation}
f(m)= f_0\, m \quad\quad {\rm where}\quad m=0,1,2,\ldots, N_s\, ,
\label{fm.def}
\end{equation}
where $N_s$ is the total number of steps. 
We restrict ourself in the present work to the zipped phase $\mu<1$,
because the simple shape of the distribution $P_N(L)\to (1-\mu)\, mu^L$ (as $N\to \infty$) in this phase
allows us to perform all computations analytically. Therefore,
the final force
$f(N_s)=f_0\, N_s$ stays below the critical force $f_c= \ln(1+e^{\beta J})/(2\beta)$ as given by Eq.~(\ref{fc_T.1}).

To define the work appropriately, consider the following general situation.
Suppose we have a system with a Hamiltonian that depends on the local degrees of freedom such as the
spins $\{s_i\}$ on a lattice and also contains a parameter $\lambda(t)$ that
evolves in continuous time (for example, the external magnetic field).
We write this Hamiltonian as $H(\{s_i\}, \lambda(t))$.
Now, we evolve the system up to a final time $t_s$ following $\lambda(t)$.
Then, quite generally, the work
done to the system up to the final time $t_s$ is defined as~\cite{jarzynski1997}
\begin{equation}
W\equiv \int_0^{t_s} dt\, {\dot \lambda}(t)\, \frac{\partial H(\{s_i\},\lambda(t))}{\partial 
\lambda(t)}\, .
\label{work_def.1}
\end{equation}
For a fixed set of spins $\{s_i\}$, as $\lambda$ evolves, the Hamiltonian 
changes, i.e., the energy changes. Hence, when integrated up to $t_s$ as
in Eq.~(\ref{work_def.1}), this gives the total energy pumped into
or released from the system, 
defined as work, just due to the change of the parameter,
and not due to spin fluctuations.

We now adapt the general definition of work in Eq.~(\ref{work_def.1}) to our zipper
model.
Here, the external force $f$ in the Hamiltonian in Eq.~(\ref{energy.1})
plays the role of the parameter $\lambda(t)$, but we assume that our evolution
occurs in discrete steps and not in continuous time. Then from Eq.~(\ref{fm.def}),
we get the discrete-time analogue of ${\dot \lambda}(t)$ in Eq.~(\ref{work_def.1}), namely
${\dot \lambda}(t)\equiv f_0$. Using furthermore $\partial_{\lambda} H \equiv -2\, L_m$
where $L_m$ is the random variable describing the free length
of the chain after the $m$-th step, we
get the discrete-time equivalent of Eq.~(\ref{work_def.1})
\begin{equation}
W= -2\, f_0\, \sum_{m=0}^{N_s-1} L_m\, .
\label{W_def}
\end{equation}
We further assume that the system equilibrates after each step (quasistatic), i.e.,
at the $m$-th step, the probability distribution of a configuration is given
by the Boltzmann weight $\propto e^{-\beta\, E_m (M,L)}$ with inverse temperature $\beta$
and with the energy function
\begin{equation}
E_m(M,L_m)= - J\, M - 2\, f(m)\, L_m\, .
\label{Em.1}
\end{equation}
It is convenient to define the rescaled work 
\begin{equation}
w= -\frac{W}{2\, f_0}= \sum_{m=0}^{N_s-1} L_m \ge 0\, .
\label{work_rescaled.1}
\end{equation}

Our goal is to find the distribution of the random variable $w$ in Eq.~(\ref{work_rescaled.1}).
To compute this, we will use the fact that $L_m$ at the $m$-th step is distributed via
the equilibrium distribution in the zipped phase as given in Eq.~(\ref{pl_zipped.1}), i.e.,
\begin{equation}
P(L_m)= (1-\mu_m)\, \mu_m^{L_m}\, ,  \quad {\rm where} \quad \mu_m= 
\frac{e^{2\, \beta\, f_0\, m}}{1+ e^{\beta\, J}}\, .
\label{PLm_def.1}
\end{equation}
Note that we have already taken the thermodynamic limit $N\to \infty$. A consequence
of Eq.~(\ref{PLm_def.1}) is
\begin{equation}
\langle z^{L_m}\rangle= \sum_{L_m=0}^{\infty} z^{L_m}\, P(L_m)= 
\frac{1-\mu_m}{1-\mu_m\,z} \, .  
\label{L_m_genf.0}
\end{equation}
We now consider the generating function of the rescaled work $w$ in Eq.~(\ref{work_rescaled.1})
and use the fact that the $L_m$'s for different $m$'s are statistically independent. This gives,
using the result in Eq.~(\ref{L_m_genf.0}), a rather nice and simple expression
\begin{equation}
\langle z^w\rangle= \prod_{m=0}^{N_s-1} \langle z^{L_m}\rangle=  
\prod_{m=0}^{N_s-1} \frac{1-\mu_m}{1-\mu_m\,z}\, ,
\label{w_genf.1}
\end{equation}
where we recall that $\mu_m$ is given in Eq.~(\ref{PLm_def.1}).

From the generating function in Eq.~(\ref{w_genf.1}), one can easily compute all
the moments and cumulants, by taking derivatives with respect to $z$ and setting $z=1$.
For this purpose, it is convenient to write $z=e^{-s}$ and derive the cumulants
by taking derivatives with respect to $s$ and set $s=0$.
Thus Eq.~(\ref{w_genf.1}) reads, in the variable $s$,
\begin{equation}
\langle e^{-s\,w}\rangle= 
\prod_{m=0}^{N_s-1} \frac{(1-\mu_m)}{\left[1-\mu_m\, e^{-s}\right]}\,.
\label{lap.1}
\end{equation} 
Consequently, the cumulant generating function is given by
\begin{equation}
\ln \left[\langle e^{-s\,w}\rangle\right]= \sum_{n=1}^{\infty} \kappa_n \frac{(-s)^n}{n!}\, ,
\label{cumul.1}
\end{equation}
where $\kappa_n$ is the $n$-th cumulant. Taking the logarithm of 
Eq.~(\ref{lap.1}) and expanding for small $s$, one can obtain all
the cumulants. For example the first two cumulants
are given by
\begin{eqnarray}
\kappa_1 &=&\langle w\rangle =  
\sum_{m=0}^{N_s-1} \frac{\mu_m}{1-\mu_m} \label{mean.1} \\
\kappa_2 &=& \langle w^2\rangle - \langle w\rangle^2 =
\sum_{n=0}^{N_s-1} \frac{\mu_m}{(1-\mu_m)^2}\, . 
\label{variance.1}
\end{eqnarray}

Let us now turn to the generating function of the full distribution
in Eq.~(\ref{lap.1}). First, we want to point out an interesting fact.
Substituting $s=-2f_0\beta$ in Eq.~(\ref{lap.1}),
we get
\begin{multline}
\langle e^{-\beta\,W}\rangle=\langle e^{2\beta f_0\,w}\rangle=
\prod_{m=0}^{N_s-1} \frac{(1-\mu_m)}{\left[1-\mu_m\, e^{2\beta f_0}\right]}\\ = 
\prod_{m=0}^{N_s-1} \frac{1+e^{\beta\, J}- e^{2\,\beta\, f_0\, m}}{1+ e^{\beta\, J}- 
e^{2\,\beta\, f_0\, (m+1)}}\, .
\label{lap_jar.1}
\end{multline}
However, now the cross terms in the numerator and denominator of Eq.~(\ref{lap_jar.1})
cancel telescopically leaving behind
\begin{equation}
\langle e^{-\beta\,W}\rangle=\frac{e^{\beta\, J}}{1+ e^{\beta\, J}- 
e^{2\,\beta\, f_0\, N_s}}\, .
\label{lap_jar.2}
\end{equation}
However, in the zipped phase where $\mu< 1$ and it is then easy to see that in the thermodynamic limit
the partition function in Eq.~(\ref{pf}) reduces, with $f=f(m)$, to
\begin{multline}
Z_N(\beta,\, f(m))\xrightarrow[N\to \infty]{} \\
\frac{(1+e^{\beta\, J})^{N+1}}{(1+e^{-\beta\, J})\left[1+e^{\beta\, J} - e^{2\, \beta\, f_0\, m}\right]}\, .
\label{pf_zipped}
\end{multline}
Hence Eq.~(\ref{lap_jar.2}) can then be expressed as
\begin{multline}
\langle e^{-\beta\,W}\rangle = \frac{Z_{N\to \infty}(\beta, \, f\,N_s)}{Z_{N\to 
\infty}(\beta, \, f(0))}= \\
\exp\left[ -\beta\, \left({\cal F}(\beta, f(m=N_s)) \right.\right.\\
  - \left.\left.{\cal F}(\beta,f(m=0))\right)\right]\, ,
\label{lap_jar.3}
\end{multline}
where ${\cal F}(\beta,\, f(m))= - (1/\beta)\, \ln Z(\beta,\, f(m))$ is the
free energy of the system at the $m$-th step.
In fact, Eq.~(\ref{lap_jar.3}) is nothing but the discrete version of the Jarzynski equality~\cite{jarzynski1997}.
 
\begin{figure}[ht]
	\includegraphics[width=0.99\columnwidth]{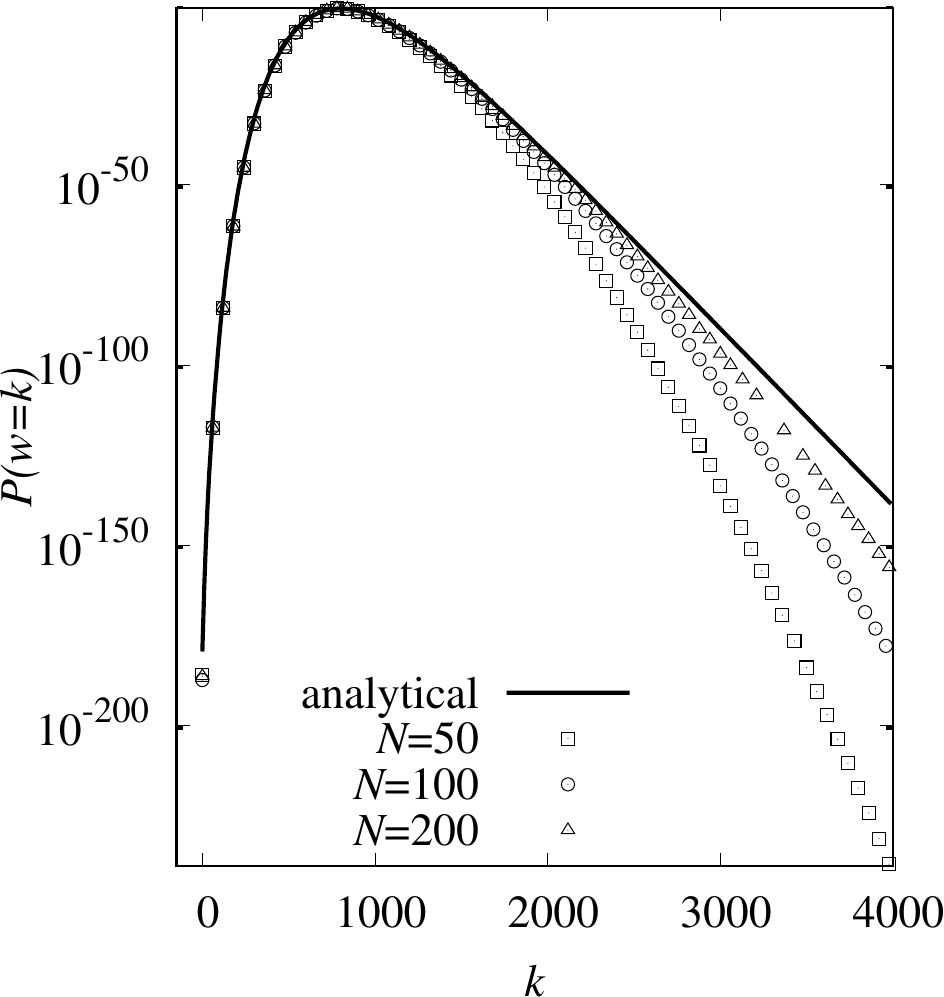}
	\caption{\label{fig:comparision:work:distribution}
          Analytical $N\to\infty$ work distribution (solid line)
          in Eq.~\eqref{wdist.1},
          measured in terms of integers $k=-W/(2f_0)$,
           for the case $\beta=J=1$
           and for $N_s=502$. 
           The symbols denote
          results obtained from numerical large-deviation sampling,
          as described in Sec.~\ref{num:large_deviation}, for
          three different chain lengths $N$.
          Errorbars are smaller than symbol size. }
\end{figure}

Note that the Jarzynski equality holds only when we set $s=-2\beta f_0$. It does not help us to compute the
full work distribution. To compute this, we need to keep a general $s$ in Eq.~(\ref{lap.1})
and try to invert this generating function.
In fact, since $w$ in Eq.~(\ref{work_rescaled.1})
is an integer, we first rewrite Eq.~(\ref{w_genf.1}) as
\begin{equation}
\langle z^k\rangle= \sum_{k=0}^{\infty} {\rm Prob.}\left[w=k|N_s\right]\, z^k=
\prod_{m=0}^{N_s-1} \frac{(1-\mu_m)}{\left[1-\mu_m\, z\right]}\, .
\label{gen_f.1}
\end{equation}
Now, one can 
formally invert the generating function in Eq.~(\ref{gen_f.1}) using
Cauchy's theorem
\begin{equation}
{\rm Prob.}\left[w=k|N_s\right]= \int_C \frac{dz}{2\pi i}\, \frac{1}{z^{k+1}}\,
\prod_{m=0}^{N_s-1} \frac{(1-\mu_m)}{\left[1-\mu_m\, z\right]}\, ,
\label{cauchy.1}
\end{equation}
where $C$ denotes a closed contour in the complex $z$ plane around $z=0$.
The integrand in Eq.~(\ref{cauchy.1}) has simple poles at $z_m= 1/\mu_m$.
Hence, one can evaluate the contour integral by computing the
residue at each pole and summing them up (with a negative sign).
This gives, after some straightforward algebra, the following explicit result
\begin{equation}
{\rm Prob.}\left[w=k|N_s\right]= \left[\prod_{l=0}^{N_s-1} (1-\mu_l)\right]\,
\sum_{m=0}^{N_s-1} \frac{\mu_m^{N_s+k}}{\prod_{n\ne m}(\mu_m-\mu_n)}\, ,
\label{wdist.1}
\end{equation}
where $\mu_m$ is given in Eq.~(\ref{PLm_def.1}). An example of the
resulting distribution is shown in Fig.~\ref{fig:comparision:work:distribution}.
Also, a comparison to numerical large-deviation data is included, which
is presented in Sec.~\ref{num:large_deviation}.

\section{The large $N_s$ behavior of the mean, the variance and the full distribution}
\label{sec:largeN_s}

In this section, we will derive the behavior of the work distribution in the scaling
limit when $N_s\to \infty$, $f_0\to 0$ but with the product 
\begin{equation}
u= 2\, \beta\, f_0\, N_s 
\label{u_def.1}
\end{equation}
fixed, corresponding to a constant final force. We consider below the mean, the variance and the full distribution separately.

\subsection{The mean}

\begin{figure}
\includegraphics[width=0.9\columnwidth]{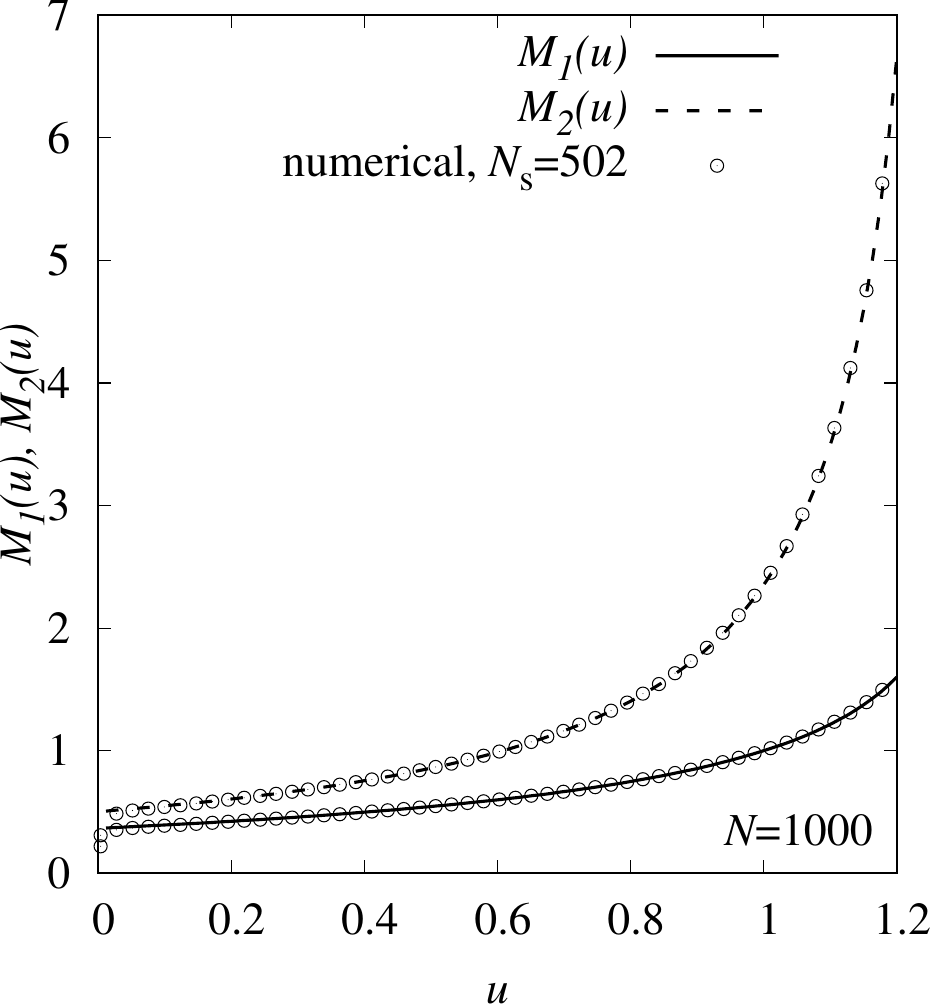}
\caption{\label{Fig:mean_variance}
  The lines show the scaling functions $M_1(u)$ for the mean work of
  Eq.~(\ref{mean_scf.1}) and $M_2(u)$ for the variance Eq.~(\ref{var_scf.1}).
 We chose $\beta=1$, $J=1$ and consequently 
$u_c=\ln(1+e^{\beta\, J})\approx 1.31$. The symbols show the numerical
results as presented in Sec.~\ref{num:large_deviation}}
\end{figure}

For the mean work $\langle w\rangle$, we have the exact formula
in Eq.~(\ref{mean.1}), namely
\begin{equation}
\langle w\rangle =
\sum_{m=0}^{N_s-1} \frac{\mu_m}{1-\mu_m}
= \sum_{m=0}^{N_s-1} \frac{e^{2\,\beta\, f_0\, m}}{1+e^{\beta J}- e^{2\,\beta\, f_0\, m}}\, ,
\label{mean.2}
\end{equation}
where we recall that $e^{2\,\beta\, f_0\, N_s}< 1+e^{\beta\, J}$.
Let us define $x= 2\, \beta\, f_0\, m$. As $m$ changes by $1$, the variable $x$
changes by $\Delta x= 2\, \beta\, f_0$. In the limit $f_0\to 0$, this change
$\Delta x\to 0$. Hence, one can replace the sum over $m$ in Eq.~(\ref{mean.2}) by an integral
over $x$ in this scaling limit. We get
\begin{multline}
\langle w\rangle \approx \frac{1}{2\, \beta\, f_0}\, \int_0^{u} dx\, 
\frac{e^x}{1+e^{\beta\, J}-e^x}= \\
\frac{1}{2\, \beta\, f_0}\, \ln\left[ 
\frac{e^{\beta \, J}}{1+e^{\beta J}- e^u}\right]\, .
\label{mean1.2}
\end{multline}
Thus, the mean work, in the scaling limit where $N_s\to \infty$, $f_0\to 0$ with
$u= 2\,\beta\, f_0\, N_s$ fixed, can be expressed in a nice scaling form
\begin{equation}
\langle w\rangle \approx N_s\, {M}_1(u=2\, \beta\, f_0\, N_s)\, ,
\label{mean_scaling.1}
\end{equation}
where the scaling function ${M}_1(u)$ is given 
for $0\le u\le u_c= \ln \left(1+ e^{\beta\, J}\right)$
by
\begin{equation}
{M}_1(u)= \frac{1}{u}\, 
\ln \left[\frac{e^{\beta \, J}}{1+e^{\beta J}- e^u}\right]\,.
\label{mean_scf.1}
\end{equation}
The scaling function ${ M}_1(u)$ is plotted in Fig.~\ref{Fig:mean_variance}
and has the following asymptotic behaviors
\begin{eqnarray}
{M}_1(u) \approx \begin{cases}
& e^{-\beta\, J}\quad {\rm as}\quad u\to 0 \, \\
\\
& -\frac{1}{u_c}\, \ln(u_c-u) \quad {\rm as}\quad u\to u_c \, .
\end{cases}
\label{M1u_asymp}
\end{eqnarray}
Thus the mean diverges very slowly (logarithmically) as $u\to u_c$ from below.

\vspace{0.3cm}

\noindent {\bf Physical significance.} The scaling behavior of the mean work in Eqs. (\ref{mean_scaling.1}) and (\ref{mean_scf.1}) has
an interesting physical significance. In fact, for this model in the thermodynamics 
limit $N\to \infty$, the partition function at step $m$ where $f_m= f_0\, m$ is given, by Eq.~(\ref{pf_zipped}).
Let us now consider the ratio
\begin{equation}
\frac{Z_{N\to \infty}(\beta, f(m=N_s))}{Z_{N\to \infty}(\beta, f(m=0))}
= \frac{e^{\beta \, J}}{1+ e^{\beta\, J}- e^{2\, \beta\, f_0\, N_s}}\, ,
\label{pf_ratio.1}
\end{equation}
where $f(m=0)=0$ and we
used Eq.~(\ref{pf_zipped}). Now consider the scaling limit $N_s\to \infty$,
$f_0\to 0$,  but with the product $u= 2\, \beta\, f_0\, N_s$ kept fixed. Then the
ratio in Eq.~(\ref{pf_ratio.1}) reduces to
\begin{equation}
\frac{Z_{N\to \infty}(\beta, f(m=N_s))}{Z_{N\to \infty}(\beta, f(m=0))}
= \frac{e^{\beta \, J}}{1+ e^{\beta\, J}- e^{u}}\, ,
\label{pf_ratio.2}
\end{equation} 
Now, from Eqs. (\ref{mean_scaling.1}) and (\ref{mean_scf.1}), we have the mean work (unscaled)
\begin{equation}
\langle W\rangle = -2\, f_0\, \langle w\rangle= -\frac{1}{\beta}\, 
\ln \left[\frac{e^{\beta \, J}}{1+e^{\beta J}- e^u}\right]\, .
\label{MW.1}
\end{equation}
Now taking logarithm on both sides of Eq.~(\ref{pf_ratio.2}), we can express the right hand side of Eq.~(\ref{MW.1}) as
\begin{multline}
\langle W\rangle = -\frac{1}{\beta}\, 
\ln\left[\frac{Z_{N\to \infty}(\beta, f(m=N_s))}{Z_{N\to \infty}(\beta, f(m=0))}
\right] \\
= {\cal F}(\beta, f(m=N_s))-{\cal F}(\beta,f(m=0))\, .
\label{MW_FE.1}
\end{multline}
Thus the mean work is exactly the free energy difference between the final equilibrium
state and the initial equilibrium state, which is expected for quasistatic evolution.

\subsection{The variance}
Also for the variance, we have the exact formula in Eq.~(\ref{variance.1}) that reads
\begin{multline}
{\rm Var}(w)=\langle w^2\rangle-\langle w\rangle^2= 
\sum_{m=0}^{N_s-1} \frac{\mu_m}{(1-\mu_m)^2}\\
= (1+e^{\beta\, J})\, 
\sum_{m=0}^{N_s-1} \frac{e^{2\,\beta\, f_0\, m}}{\left[1+e^{\beta\, J}- 
e^{2\,\beta\, f_0\, m}\right]^2}\, .
\label{var.2}
\end{multline}
As in the case of the mean, we define the variable $x= 2\, \beta\, f_0\, m$ which
becomes continuous in the $f_0\to 0$ limit. Hence, in the scaling limit in 
Eq.~(\ref{u_def.1}), we can replace the sum by an integral and write
\begin{equation}
{\rm Var}(w)\approx N_s\, (1+e^{\beta\, J})\, \frac{1}{u}\, \int_0^{u} dx\, 
\frac{e^x}{\left[1+e^{\beta\, J}-e^x\right]^2}\, .
\label{var.3}
\end{equation}
Performing the integral explicitly, we then have the scaling behavior
\begin{equation}
{\rm Var}(w)\approx N_s\, {M}_2(u= 2\,\beta\, f_0\, N_s)\, ,
\label{var_scaling.1}
\end{equation}
where the scaling function ${M}_2(u)$ is 
for $0\le u\le u_c= \ln \left(1+ e^{\beta\, J}\right)$
given explicitly by
\begin{equation}
{M}_2(u)= (1+ e^{-\beta\, J}) \frac{(e^{u}-1)}{u\, (1+e^{\beta\, J}-e^u)}\, .
\label{var_scf.1}
\end{equation}
The scaling function ${M}_2(u)$ is plotted in Fig.~\ref{Fig:mean_variance}
and has the following asymptotic behaviors
\begin{eqnarray}
{M}_2(u) \approx \begin{cases}
& e^{-\beta\, J} (1+ e^{-\beta\, J})\quad {\rm as}\quad u\to 0 \, \\
\\
& \frac{1}{u_c\, (u_c-u)} \quad {\rm as}\quad u\to u_c \, .
\end{cases}
\label{M2u_asymp}
\end{eqnarray}
Thus as $u\to u_c$ from below, the variance diverges as $1/(u_c-u)$,
which is faster as compared to the
mean in Eq.~(\ref{M1u_asymp}).

\subsection{The full work distribution}
Here, we will consider the full work distribution in the scaling limit $N_s\to \infty$,
$f_0\to 0$ with their product $u= 2\, \beta\, f_0\, N_s$ kept fixed. Our starting point
is the exact Cauchy representation of the work distribution in Eq.~(\ref{cauchy.1}), which
can be re-written as
\begin{multline}
{\rm Prob.}\left[w=k|N_s\right]= \\
A\, \int_C \frac{dz}{2\pi i}\, \exp\left[-(k+1)\, \ln(z)
- \sum_{m=0}^{N_s-1} \ln(1- \mu_m\, z)\right]\,,
\label{cauchy.2}
\end{multline}
where   $A= \prod_{m=0}^{N_s-1} (1-\mu_m)$, 
and we recall that $\mu_m$ is given by Eq.~(\ref{PLm_def.1}).

\begin{figure}[h]
	\includegraphics[width=0.99\columnwidth]{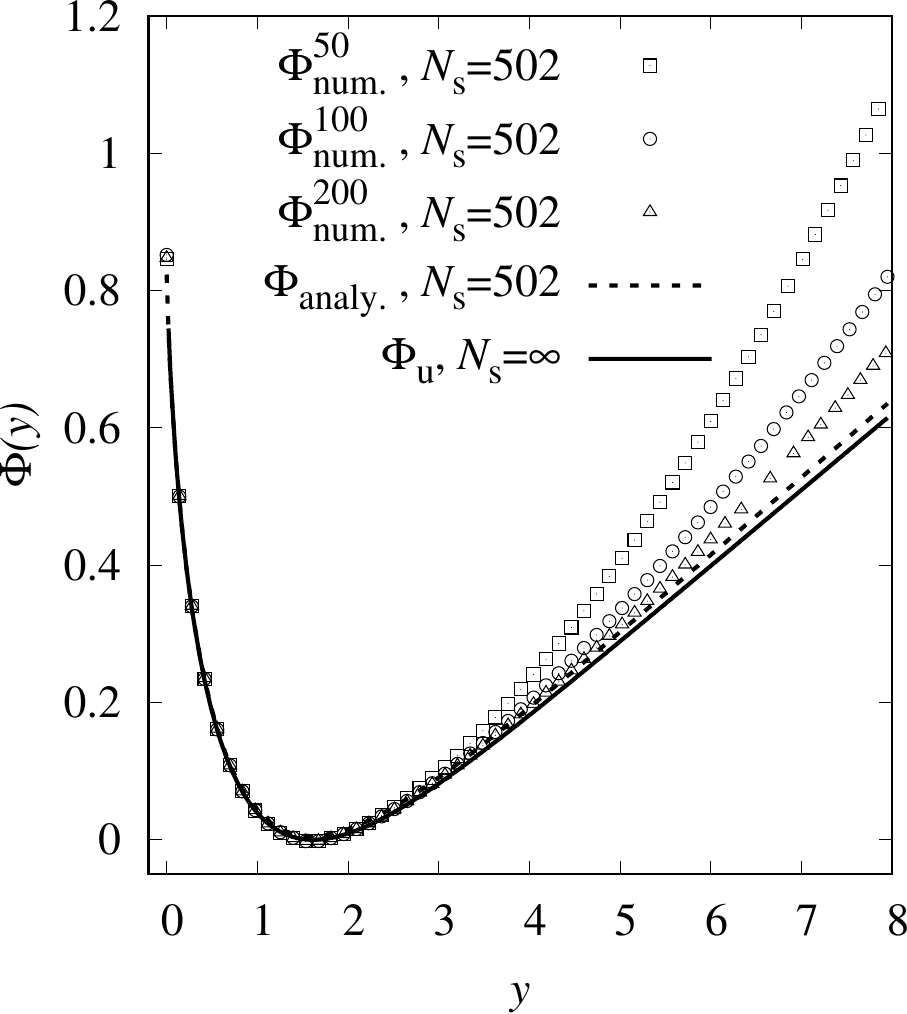}
	\caption{\label{fig:comparison:rate:function}
          Analytical (line) $N_s\to\infty$
           rate function $\Phi_u(y)$ from Eq.~(\ref{ratef.2})
           for fixed $u= 1.2$.
           The rate function has a minimum at
            $y_{\rm min}= M_1(u=1.2)\approx 1.6$
           and has the quadratic behavior near the minimum as in
            Eq.~(\ref{quadratic.1}).
           Also shown is the analytical result for finite $N_s=502$ (broken
            line)
           and numerical estimates $\Phi^N_{\text{num.}}(y)$
           as obtained from large-deviation simulations, see
           Sec.~\ref{num:large_deviation}.
           Errorbars are smaller than symbol size.}
\end{figure}

To proceed we set $k= y\, N_s$ where $y$ is of $O(1)$ in the large $N_s$ limit.
Taking the scaling limit as usual, i.e., by defining $x= 2\,\beta\, f_0\,m$
and replacing the sum inside the exponential by an integral, one finds after straightforward
algebra, for  $y=\frac{k}{N_s}$ and $u=  2\,\beta\, f_0\, N_s$,
\begin{multline}
{\rm Prob.}\left[w=k|N_s\right] \\
\approx A\,\, e^{N_s\, \ln (1+e^{\beta\,J})}\,
\int_C \frac{dz}{2\pi i}\, \exp\left[- N_s\, S\left(z|y, u
\right)\right]\, ,
\label{cauchy_sc.1}
\end{multline}  
where the action $S$ is given explicitly by
\begin{equation}
S\left(z | y, u\right)= y\, \ln z + \frac{1}{u}\, \int_0^u \ln 
\left(1+e^{\beta\, J}- z\, e^x\right)\, dx\, .
\label{action.1}
\end{equation}
The idea is now to perform a saddle point analysis of the integral in the large $N_s$
limit. Taking derivative with respect to $z$ and setting
\begin{equation}
\partial_z S(z|y,u)=0 \, ,
\label{saddle.0}
\end{equation}
gives the saddle point explicitly
\begin{equation}
z^*(y|u)= \frac{(1+e^{\beta\, J})\,(e^{u\, y}-1)}{\left(e^{u\,(y+1)}-1\right)}\, .
\label{saddle.1}
\end{equation}
Similarly the prefactor $A\,\, e^{N_s\, \ln (1+e^{\beta\,J})}$ in 
Eq.~(\ref{cauchy_sc.1}) can be analyzed in the large $N_s$ limit giving
\begin{equation}
A\,\, e^{N_s\, \ln (1+e^{\beta\,J})}\approx \exp\left[N_s\, \frac{1}{u}\, \int_0^u \ln \left(
1+ e^{\beta\, J}- e^x\right)\, dx\right] \, .
\label{prefactor.1}
\end{equation}
Evaluating the integral by the saddle point method for large $N_s$
and using the result in Eq.~(\ref{prefactor.1}), the work distribution
can be expressed in a nice large-deviation form (for fixed $u=2\,\beta\, f_0\, N_s$
and $N_s$ large)
\begin{equation}
{\rm Prob.}\left[w=k|N_s\right] \approx \exp\left[- N_s\, \Phi_u\left(\frac{k}{N_s}=y
\right)\right]\, ,
\label{ldf.1}
\end{equation}
where the rate function $\Phi_u(y)$ is given by
\begin{multline}
\Phi_u(y)= y\, \ln \left(z^*(y|u)\right) \\
+\frac{1}{u}\, \int_0^u \ln \left[ 
\frac{1+e^{\beta\, J}- z^*(y|u)\, e^x}{1+ e^{\beta\, J}- e^x}\right]\, dx\, ,
\label{ratef.1}
\end{multline}
with $z^*(y|u)$ given in Eq.~(\ref{saddle.1}). Performing the 
integral explicitly,  we get
\begin{multline}
\Phi_u(y)= y\, \ln \left(z^*(y|u)\right)\\
+ \frac{1}{u}\, \bigg[
{\rm Li}_2\left(\frac{z^*(y|u)}{1+e^{\beta\, J}}\right)
-{\rm Li}_2\left(\frac{1}{1+e^{\beta\, J}}\right) \\
+{\rm Li}_2\left(\frac{e^{u}}{1+e^{\beta\, J}}\right)
-{\rm Li}_2\left(\frac{e^{u}\, z^*(y|u)}{1+e^{\beta\, J}}\right) 
\bigg]\, , 
\label{ratef.2}
\end{multline}
where 
\begin{equation}
{\rm Li}_2(z)= \sum_{n=1}^{\infty}\frac{z^n}{n^2}\, ,
\label{dilog.1}
\end{equation}
is the dilogarithm function.
A plot of the rate function $\Phi_u(y)$ vs. $y$ (for fixed $u$) is given in
Fig.~\ref{fig:comparison:rate:function}.
By taking the derivative with respect to $y$ (for fixed $u$) and setting $\partial_y \Phi_u(y)=0$
gives the value $y_{\rm min}(u)$ where the rate function $\Phi_u(y)$ has its minimum.
It is not difficult to show that
\begin{equation}
y_{\rm min}(u)= \frac{1}{u}\, \ln \left[\frac{e^{\beta \, J}}{1+e^{\beta J}- e^u}\right]
\equiv { M}_1(u)\, ,
\label{ymin.1}
\end{equation}
where ${ M}_1(u)$ defined in Eq.~(\ref{mean_scaling.1}) is just $\langle w\rangle/N_s$.
Indeed, one expects the rate function $\Phi_u(y)$ to be a convex function with
a minimum at $y=y_{\rm min}(u)={ M}_1(u)$ and must have a quadratic form
near this minimum
\begin{equation}
\Phi_u(y) \approx \frac{\left(y- { M}_1(u)\right)^2}{2\, M_2(u)}\, ,
\label{quadratic.1}
\end{equation}
where $M_2(u)$ is the scaling function associated to the variance 
in Eq.~(\ref{var_scaling.1}) and computed explicitly in Eq.~(\ref{var_scf.1}).
Indeed, by expanding $\Phi_u(y)$ up to quadratic order around $y=y_{\rm min}(u)$,
one does recover ${M}_2(u)$ from the rate function. The asymptotic behaviors
of $\Phi_u(y)$ as $y\to 0$ and $y\to \infty$ can also be deduced easily.
Let us define the parameter
\begin{equation}
a= 1+ e^{\beta\, J}\, .
\label{def_a}
\end{equation}
In terms of this parameter $a$, we can express the small and large $y$ asymptotics
of $\Phi_u(y)$ (for fixed $u$) as follows
\begin{eqnarray}
\Phi_u(y) = \begin{cases}
& \frac{1}{u}\, \left({\rm Li}_2\left(\frac{e^{u}}{a}\right)-{\rm Li}_2\left(\frac{1}{a}\right)
\right) + \\
&  \quad \quad y\, \ln y +O(y)\, , \quad {\rm as}\quad y\to 0\, \\
\\
& \frac{1}{u}\left(-{\rm Li}_2\left(\frac{1}{a}\right)
+{\rm Li}_2\left(\frac{e^{u}}{a}\right) +{\rm Li}_2\left(e^{-u}\right)-\frac{\pi^2}{6}\right)\\
& +\ln(a\, e^{-u})\, y + O\left(\frac{1}{y}\right)\,
 \quad {\rm as}\quad y\to \infty\, .
\end{cases}
\label{phi_asymp}
\end{eqnarray} 
Interestingly $\Phi_u(0)=\left[{\rm Li}_2\left(\frac{e^{u}}{a}\right)-{\rm Li}_2\left(\frac{1}{a}\right)\right]/u$ is a 
positive constant. This implies, from Eq.~(\ref{ldf.1}), that the probability of vanishing work, i.e.,
$w<< N_s$ decays exponentially with increasing $N_s$ as
\begin{multline}
{\rm Prob.}\left[w=k|N_s\right] \xrightarrow[k\ll N_s]{} \exp\left[- \theta(u)\, N_s\right]\, \quad \\
{\rm where}\quad \theta(u)= \Phi_u(0)= \frac{1}{u}\, \left({\rm Li}_2\left(\frac{e^{u}}{a}\right)-{\rm Li}_2\left(\frac{1}{a}\right)
\right)\, .
\label{zero_work.1}
\end{multline}
Also, since $\Phi_u(y)$ increases linearly with $y$ for large $y$,
see the second line
of Eq.~(\ref{phi_asymp}), it follows again from Eq.~(\ref{ldf.1}) that the probability
of a very large work $w=k>> N_s$ becomes independent of $N_s$ and decays exponentially
with increasing $k$
\begin{equation}
{\rm Prob.}\left[w=k|N_s\right] \xrightarrow[k\gg N_s]{} \exp\left[- k\, \ln (a\, e^u)\right]=
\frac{1}{\left(a\, e^{u}\right)^k}\, .
\label{large_work.1}
\end{equation}

\section{Large-Deviation Simulations\label{num:large_deviation}}

Complementary to the analytical calculations of the work distribution, numerical
simulations were performed, especially to study finite-size effects.

\begin{figure}
\begin{center}
\includegraphics[width=0.8\columnwidth]{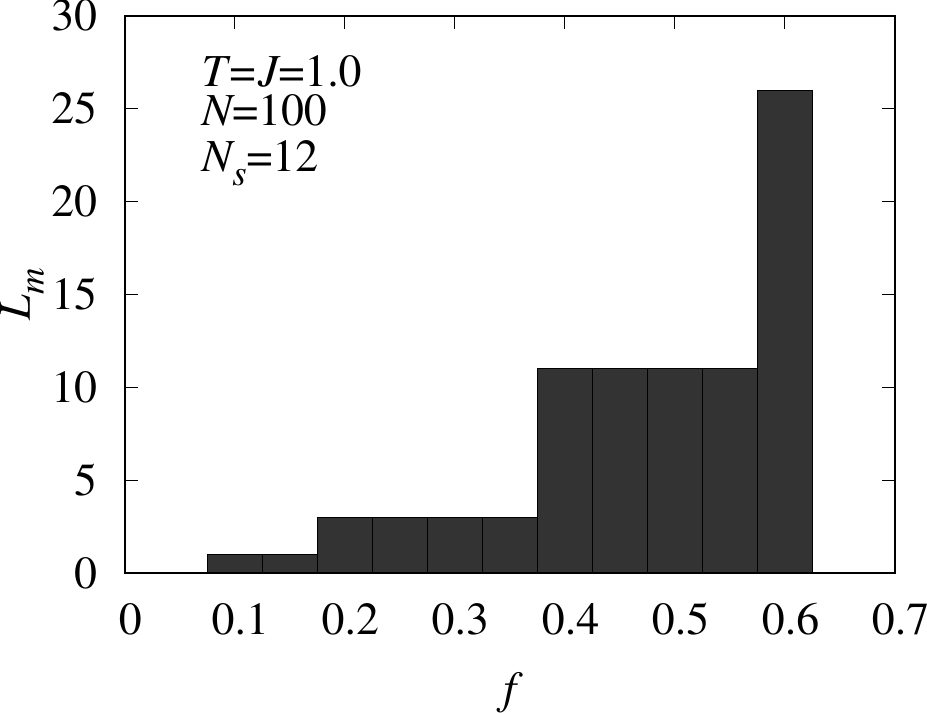}
\end{center}
\caption{A sample force-extension curve where the free length $L_m$ at step
  $m$ is shown as function of the force $f$ for a single random realization
  of the unzipping
  process, here for length $N=100$, $N_s$=12 steps and parameters $T=J=1$.
  The area under the curve is the work, times a factor of -2 by
  Eq.~(\ref{work_def.1}). 
\label{fig:sample_L_f}}
\end{figure}

The work simulation works as follows:
The linear protocol is discretized into, here, $N_s=502$ points resulting in force values $f_m=m f_0,\; m\in[0,1,2, \dots, N_s]$, where $f_0$ is a constant chosen such that the final force is $N_s f_0 = 0.6$, which is smaller than the
critical force $f_c \approx 0.6566$ for the case $\beta=J=1$.
Therefore the phase transition should not contribute significantly.
At each force value a sample from the equilibrium distribution is drawn using the algorithm as described Sec.~\ref{sec:sampling},
yielding a total of $N_s$ configurations where each one exhibits
a free length $L_m$, i.e., seen from the
beginning (left), a corresponding number $L_m$ of 0's before the first
occurrence of a 1. A sample of such a \emph{force-extension} curve is shown in
Fig.~\ref{fig:sample_L_f}.
The work of an entire process is then given Eq.~(\ref{W_def}),
which is twice the area under the curve in the figure.
We considered the chain lengths
$N\in \left\lbrace 50, 100, 200 \right\rbrace$.

By running the process many times, for the desired values of $f_0$ and $N_s$
one can obtain histograms of the work in the high-probability
region. This allowed us to measure directly the mean and the variance
of the work. The resulting rescaled work and variance functions $M_1(u)$
and $M_2(u)$, see Eqs.~(\ref{mean_scaling.1}) and (\ref{var_scaling.1}),
are shown in Fig.~\ref{Fig:mean_variance}. An almost perfect agreement
of analytical results (for $N_s=\infty$) and numerical results is visible,
except for very small values of the rescaled force $u$, where the finiteness
of the number of steps and sequence length becomes slightly visible.


For a meaningful comparison of the full
distribution to the analytical result, it is necessary
to resolve the work distribution over a large range of support,
down to probabilities as small as $10^{-200}$. For this purpose, 
a large-deviation algorithm \cite{work_ising2014} was employed.
Note that this approach was already used to measure
work distributions for unzipping processes of a more sophisticated RNA
model \cite{Werner_2021}.
The method's basic
idea is to treat the underlying random numbers ${\bf \xi}=(\xi_1,\ldots,\xi_K)$, required by
a single simulation of one full
work process, as a state variable in a Markov-chain Monte Carlo process
${\bf \xi}^{(1)} \to {\bf \xi}^{(2)} \to \ldots$ 
with a target distribution that has an additional bias proportional
to $\exp{(W / \theta)}$. Here $\theta\in [-0.0005, \dots, 0.0005]$ is a temperature-like parameter allowing to control the regime in which work values are predominantly generated.
Since every equilibrium sample needs $N$ random numbers and each work process requires $N_s$ samples at different force values, the state vector ${\bf \xi}$
of random numbers has a size of $K=N \times N_s$. Each unzipping process
is fully determined by ${\bf \xi}$, therefore the work is a deterministic
function $W=W({\bf \xi})$.

One Monte-Carlo step in the Markov chain of random number vectors
consists of
\begin{itemize}
\item randomly selecting some entries of the current
  state vector ${\bf \xi}^{(t)}$ and redrawing them, resulting
  in a trial state ${\bf \xi}'$.
\item performing one  full work simulation with 
this trial state of random numbers
\item obtaining the corresponding work $W'=W({\bf \xi}')$
\item finally accepting or rejecting the trial state with
  Metropolis probability 
  $p_{\mathrm{Metr.}}=\min{\left\lbrace 1, \exp{(\Delta W / \theta)}\right\rbrace }$, where $\Delta W=W'-W({\bf \xi}^{(t)})$
  is the work difference between current and trial state.
\end{itemize}
Equilibration was ensured as follows: one starts the Markov chain 
with random as well as with extreme vectors of random numbers, such that the
corresponding initial work values are very different, respectively. If
for these very different initial conditions after a while the work values
agree withing fluctuations, the chain can be considered as
equilibrated.
For more details see Ref.~\cite{largest-2011}.
We considered 90 different values of $\theta$.
For each one, a separate MC simulation is performed with at most $3.2 \times 10^8$
MC-steps.
This has yielded up to 40000 sample points each after the initial 
equilibration phase and correlations were removed from the chain of work 
values.
Hence, for each value of $\theta$ a
distribution $P_\theta(W)$ is obtained. These are then combined to the final overall work distribution via the Ferrenberg-Swendsen method 
\cite{Ferrenberg_1989} using a convenient tool \cite{werner2022}.

Fig.~\ref{fig:comparision:work:distribution} shows the exact analytical and numerically estimated work distributions for the rescaled work $w=k=-W/2f_0$. 
The distributions could be obtained
down to probability values as small as $10^{-200}$.
For $k<1250$, they show good agreement for all simulated system sizes.
Finite-size effect become relevant in the regime $k>1250$.
Nevertheless, the curves tend toward the analytical distribution as the system
size increases.

Due to the quasistatic nature, i.e., sampling in perfect thermal equilibrium,
each work value is a sum of $N_s$ statistical independent values $L_m$. 
But these values are not identically distributed due to the different 
force parameter values at each of the process
steps. This explains why the conditions
of the central limit theorem are not fulfilled. The resulting distributions 
therefore do not need to be Gaussian, which they apparently are not.

An estimate of the rate functions from the numerical determined work
distribution $P^N_{\text{num.}}$ is given by:
\begin{equation}\label{eq:rate:function:estimate}
\Phi^N_{\text{num.}}(y) = -\frac{1}{N_s}\ln(P^N_{\text{num.}}(k=y N_s))\,.
\end{equation}
Since the analytical rate function in eq. \eqref{ratef.2} considers the case $N_s \rightarrow\infty$, it is also worth looking at
\begin{equation}\label{eq:rate:function:estimate:analytical}
\Phi_{\text{analy.}}(y) = -\frac{1}{N_s}\ln(\text{Prob.}[w=y N_s|N_s])\, ,
\end{equation}
determined from the exact work distribution but for finite $N_s$ in eq. \eqref{wdist.1}.

The different rate functions are displayed in
Fig.~\ref{fig:comparison:rate:function}.
Again, the numerical curves tend towards the analytical rate functions with increasing system size, where they are matching each other for $y<3$.
Finite-size effects become relevant for $y>3$.
The analytical rate function for finite $N_s$ is only slightly different 
from the $N_s\to\infty$ limiting one towards
higher values of $y$.
This indicates that the influence of a finite number of protocol points $N_s$
is minor compared to that of a finite system size $N$.

\section{Summary and Discussion}
\label{Conclusion}

In summary, we have calculated analytically and verified numerically the distribution
$P(W)$ of work performed in unzipping an infinitely long closed hairpin structure under an external force,
applied quasistatically in the zipped phase.
This model has many interacting degrees of freedom
and goes beyond past analytical works where either single-particle
systems or models with simple, i.e., harmonic, interactions where
considered.

As one of the main steps leading to this calculation of the work distribution, we computed exactly the full equilibrium 
distribution $P_N(L)$ of the free length $L$ of an infinitely long hairpin ($N\to \infty$ limit), thus going beyond previous studies that focused
mainly on the average $\langle L\rangle$
as a function of system parameters. We find that the large-$N$ distribution $P_{N\to \infty}(L)$
drastically changes its shape at the
unzipping transition point $f=f_c$, and is, in particular, very broad right at the
transition.

Based on this result, we were able to compute $P(W)$ in the zipped phase analytically for all $W$.
It also allowed us to evaluate $P(W)$ numerically
over its full range of support, resulting in probabilities as small
as $10^{-200}$ for the selected values parameters.
We find generally a very good agreement
between analytical and numerical approaches, 
except expected finite-size effects which are
present in the numerical simulations. There is also a dependence
on the number of steps of changes of the force, but here the
influence on the results is rather limited.

In this paper, we have restricted our calculation of $P(W)$ only in the zipped phase, for
simplicity. It would be interesting to extend our calculation that encompasses both phases.
Finally, our results are valid for quasistatic processes. It would also
be interesting to study the work distribution for a genuinely
far from equilibrium process. 
One could describe
them, e.g., by considering the distribution
$P(L,t)$, $t$ being a time or step counter,
and describing the dynamics by allowing for transitions between
neighboring states.
Still, for this model,
the part beyond the first pair is ``shielded'' from the external
force, i.e., is always in equilibrium. Only the free part
couples to the external force and is influenced by the rate of
force change. Thus, one could expect an only small influence
of the speed of the process, but this remains to be verified in future work.

\begin{acknowledgments}
  SNM acknowledges the Alexander von Humboldt foundation for the
  Gay Lussac-Humboldt
prize that allowed a visit to the Physics department at Oldenburg University, Germany where this work
initiated. He also acknowledges the kind hospitality of the Oldenburg University.
  The simulations were performed at the
  the HPC cluster ROSA, located at the University of Oldenburg
  (Germany) and
    funded by the DFG through its Major Research Instrumentation Program
    (INST 184/225-1 FUGG) and the Ministry of
    Science and Culture (MWK) of the
    Lower Saxony State. 
\end{acknowledgments}

\bibliography{references}

\begin{thebibliography}{54}%
\makeatletter
\providecommand \@ifxundefined [1]{%
 \@ifx{#1\undefined}
}%
\providecommand \@ifnum [1]{%
 \ifnum #1\expandafter \@firstoftwo
 \else \expandafter \@secondoftwo
 \fi
}%
\providecommand \@ifx [1]{%
 \ifx #1\expandafter \@firstoftwo
 \else \expandafter \@secondoftwo
 \fi
}%
\providecommand \natexlab [1]{#1}%
\providecommand \enquote  [1]{``#1''}%
\providecommand \bibnamefont  [1]{#1}%
\providecommand \bibfnamefont [1]{#1}%
\providecommand \citenamefont [1]{#1}%
\providecommand \href@noop [0]{\@secondoftwo}%
\providecommand \href [0]{\begingroup \@sanitize@url \@href}%
\providecommand \@href[1]{\@@startlink{#1}\@@href}%
\providecommand \@@href[1]{\endgroup#1\@@endlink}%
\providecommand \@sanitize@url [0]{\catcode `\\12\catcode `\$12\catcode
  `\&12\catcode `\#12\catcode `\^12\catcode `\_12\catcode `\%12\relax}%
\providecommand \@@startlink[1]{}%
\providecommand \@@endlink[0]{}%
\providecommand \url  [0]{\begingroup\@sanitize@url \@url }%
\providecommand \@url [1]{\endgroup\@href {#1}{\urlprefix }}%
\providecommand \urlprefix  [0]{URL }%
\providecommand \Eprint [0]{\href }%
\providecommand \doibase [0]{http://dx.doi.org/}%
\providecommand \selectlanguage [0]{\@gobble}%
\providecommand \bibinfo  [0]{\@secondoftwo}%
\providecommand \bibfield  [0]{\@secondoftwo}%
\providecommand \translation [1]{[#1]}%
\providecommand \BibitemOpen [0]{}%
\providecommand \bibitemStop [0]{}%
\providecommand \bibitemNoStop [0]{.\EOS\space}%
\providecommand \EOS [0]{\spacefactor3000\relax}%
\providecommand \BibitemShut  [1]{\csname bibitem#1\endcsname}%
\let\auto@bib@innerbib\@empty
\bibitem [{\citenamefont {Seifert}(2012)}]{seifert2012}%
  \BibitemOpen
  \bibfield  {author} {\bibinfo {author} {\bibfnamefont {U.}~\bibnamefont
  {Seifert}},\ }\href {\doibase 10.1088/0034-4885/75/12/126001} {\bibfield
  {journal} {\bibinfo  {journal} {Rep. Progr. Phys.}\ }\textbf {\bibinfo
  {volume} {75}} (\bibinfo {year} {2012}),\
  10.1088/0034-4885/75/12/126001}\BibitemShut {NoStop}%
\bibitem [{\citenamefont {Peliti}\ and\ \citenamefont
  {Pigolotti}(2021)}]{peliti2021}%
  \BibitemOpen
  \bibfield  {author} {\bibinfo {author} {\bibfnamefont {L.}~\bibnamefont
  {Peliti}}\ and\ \bibinfo {author} {\bibfnamefont {S.}~\bibnamefont
  {Pigolotti}},\ }\href@noop {} {\emph {\bibinfo {title} {Stochastic
  Thermodynamics: An Introduction}}}\ (\bibinfo  {publisher} {Princeton
  University Press},\ \bibinfo {address} {Princeton},\ \bibinfo {year}
  {2021})\BibitemShut {NoStop}%
\bibitem [{\citenamefont {Jarzynski}(1997)}]{jarzynski1997}%
  \BibitemOpen
  \bibfield  {author} {\bibinfo {author} {\bibfnamefont {C.}~\bibnamefont
  {Jarzynski}},\ }\href {\doibase 10.1103/PhysRevLett.78.2690} {\bibfield
  {journal} {\bibinfo  {journal} {Phys. Rev. Lett.}\ }\textbf {\bibinfo
  {volume} {78}},\ \bibinfo {pages} {2690} (\bibinfo {year}
  {1997})}\BibitemShut {NoStop}%
\bibitem [{\citenamefont {Crooks}(1998)}]{crooks1998}%
  \BibitemOpen
  \bibfield  {author} {\bibinfo {author} {\bibfnamefont {G.~E.}\ \bibnamefont
  {Crooks}},\ }\href {\doibase 10.1023/A:1023208217925} {\bibfield  {journal}
  {\bibinfo  {journal} {J. Stat.\ Phys.}\ }\textbf {\bibinfo {volume} {90}},\
  \bibinfo {pages} {1481} (\bibinfo {year} {1998})}\BibitemShut {NoStop}%
\bibitem [{\citenamefont {Collin}\ \emph {et~al.}(2005)\citenamefont {Collin},
  \citenamefont {Ritort}, \citenamefont {Jarzynski}, \citenamefont {Smith},
  \citenamefont {Tinoco},\ and\ \citenamefont {Bustamante}}]{collin2005}%
  \BibitemOpen
  \bibfield  {author} {\bibinfo {author} {\bibfnamefont {D.}~\bibnamefont
  {Collin}}, \bibinfo {author} {\bibfnamefont {F.}~\bibnamefont {Ritort}},
  \bibinfo {author} {\bibfnamefont {C.}~\bibnamefont {Jarzynski}}, \bibinfo
  {author} {\bibfnamefont {S.~B.}\ \bibnamefont {Smith}}, \bibinfo {author}
  {\bibfnamefont {I.}~\bibnamefont {Tinoco}}, \ and\ \bibinfo {author}
  {\bibfnamefont {C.}~\bibnamefont {Bustamante}},\ }\href {\doibase
  {10.1038/nature04061}} {\bibfield  {journal} {\bibinfo  {journal} {{Nature}}\
  }\textbf {\bibinfo {volume} {{437}}},\ \bibinfo {pages} {231} (\bibinfo
  {year} {{2005}})}\BibitemShut {NoStop}%
\bibitem [{\citenamefont {Ciliberto}(2017)}]{ciliberto2017}%
  \BibitemOpen
  \bibfield  {author} {\bibinfo {author} {\bibfnamefont {S.}~\bibnamefont
  {Ciliberto}},\ }\href {\doibase 10.1103/PhysRevX.7.021051} {\bibfield
  {journal} {\bibinfo  {journal} {Phys. Rev. X}\ }\textbf {\bibinfo {volume}
  {7}},\ \bibinfo {pages} {021051} (\bibinfo {year} {2017})}\BibitemShut
  {NoStop}%
\bibitem [{\citenamefont {Hartmann}(2014)}]{work_ising2014}%
  \BibitemOpen
  \bibfield  {author} {\bibinfo {author} {\bibfnamefont {A.~K.}\ \bibnamefont
  {Hartmann}},\ }\href {\doibase 10.1103/PhysRevE.89.052103} {\bibfield
  {journal} {\bibinfo  {journal} {Phys. Rev. E}\ }\textbf {\bibinfo {volume}
  {89}},\ \bibinfo {pages} {052103} (\bibinfo {year} {2014})}\BibitemShut
  {NoStop}%
\bibitem [{\citenamefont {Werner}\ and\ \citenamefont
  {Hartmann}(2021)}]{Werner_2021}%
  \BibitemOpen
  \bibfield  {author} {\bibinfo {author} {\bibfnamefont {P.}~\bibnamefont
  {Werner}}\ and\ \bibinfo {author} {\bibfnamefont {A.~K.}\ \bibnamefont
  {Hartmann}},\ }\href {\doibase 10.1103/PhysRevE.104.034407} {\bibfield
  {journal} {\bibinfo  {journal} {Phys. Rev. E}\ }\textbf {\bibinfo {volume}
  {104}},\ \bibinfo {pages} {034407} (\bibinfo {year} {2021})}\BibitemShut
  {NoStop}%
\bibitem [{\citenamefont {Speck}\ and\ \citenamefont
  {Seifert}(2004)}]{speck2004}%
  \BibitemOpen
  \bibfield  {author} {\bibinfo {author} {\bibfnamefont {T.}~\bibnamefont
  {Speck}}\ and\ \bibinfo {author} {\bibfnamefont {U.}~\bibnamefont
  {Seifert}},\ }\href {\doibase 10.1103/PhysRevE.70.066112} {\bibfield
  {journal} {\bibinfo  {journal} {Phys. Rev. E}\ }\textbf {\bibinfo {volume}
  {70}},\ \bibinfo {pages} {066112} (\bibinfo {year} {2004})}\BibitemShut
  {NoStop}%
\bibitem [{\citenamefont {Speck}\ and\ \citenamefont
  {Seifert}(2005)}]{speck2005}%
  \BibitemOpen
  \bibfield  {author} {\bibinfo {author} {\bibfnamefont {T.}~\bibnamefont
  {Speck}}\ and\ \bibinfo {author} {\bibfnamefont {U.}~\bibnamefont
  {Seifert}},\ }\href@noop {} {\bibfield  {journal} {\bibinfo  {journal} {Eur.
  Phys. J. B}\ }\textbf {\bibinfo {volume} {43}},\ \bibinfo {pages} {521–527}
  (\bibinfo {year} {2005})}\BibitemShut {NoStop}%
\bibitem [{\citenamefont {Lua}\ and\ \citenamefont {Grosberg}(2005)}]{lua2005}%
  \BibitemOpen
  \bibfield  {author} {\bibinfo {author} {\bibfnamefont {R.~C.}\ \bibnamefont
  {Lua}}\ and\ \bibinfo {author} {\bibfnamefont {A.~Y.}\ \bibnamefont
  {Grosberg}},\ }\href {\doibase 10.1021/jp0455428} {\bibfield  {journal}
  {\bibinfo  {journal} {J. Phys. Chem. B}\ }\textbf {\bibinfo {volume} {109}},\
  \bibinfo {pages} {6805} (\bibinfo {year} {2005})},\ \bibinfo {note} {pMID:
  16851766}\BibitemShut {NoStop}%
\bibitem [{\citenamefont {Quan}\ \emph {et~al.}(2008)\citenamefont {Quan},
  \citenamefont {Yang},\ and\ \citenamefont {Sun}}]{quan2008}%
  \BibitemOpen
  \bibfield  {author} {\bibinfo {author} {\bibfnamefont {H.~T.}\ \bibnamefont
  {Quan}}, \bibinfo {author} {\bibfnamefont {S.}~\bibnamefont {Yang}}, \ and\
  \bibinfo {author} {\bibfnamefont {C.~P.}\ \bibnamefont {Sun}},\ }\href
  {\doibase 10.1103/PhysRevE.78.021116} {\bibfield  {journal} {\bibinfo
  {journal} {Phys. Rev. E}\ }\textbf {\bibinfo {volume} {78}},\ \bibinfo
  {pages} {021116} (\bibinfo {year} {2008})}\BibitemShut {NoStop}%
\bibitem [{\citenamefont {Speck}(2011)}]{speck2011}%
  \BibitemOpen
  \bibfield  {author} {\bibinfo {author} {\bibfnamefont {T.}~\bibnamefont
  {Speck}},\ }\href {\doibase 10.1088/1751-8113/44/30/305001} {\bibfield
  {journal} {\bibinfo  {journal} {J. Phys. A}\ }\textbf {\bibinfo {volume}
  {44}},\ \bibinfo {pages} {305001} (\bibinfo {year} {2011})}\BibitemShut
  {NoStop}%
\bibitem [{\citenamefont {Saha}\ and\ \citenamefont
  {Mukherji}(2014)}]{saha2014}%
  \BibitemOpen
  \bibfield  {author} {\bibinfo {author} {\bibfnamefont {B.}~\bibnamefont
  {Saha}}\ and\ \bibinfo {author} {\bibfnamefont {S.}~\bibnamefont
  {Mukherji}},\ }\href {\doibase 10.1088/1742-5468/2014/08/P08014} {\bibfield
  {journal} {\bibinfo  {journal} {J. Stat. Mech.}\ }\textbf {\bibinfo {volume}
  {2014}},\ \bibinfo {pages} {P08014} (\bibinfo {year} {2014})}\BibitemShut
  {NoStop}%
\bibitem [{\citenamefont {Engel}(2009)}]{engel2009}%
  \BibitemOpen
  \bibfield  {author} {\bibinfo {author} {\bibfnamefont {A.}~\bibnamefont
  {Engel}},\ }\href {\doibase 10.1103/PhysRevE.80.021120} {\bibfield  {journal}
  {\bibinfo  {journal} {Phys. Rev. E}\ }\textbf {\bibinfo {volume} {80}},\
  \bibinfo {pages} {021120} (\bibinfo {year} {2009})}\BibitemShut {NoStop}%
\bibitem [{\citenamefont {Nickelsen}\ and\ \citenamefont
  {Engel}(2011)}]{nickelsen2011}%
  \BibitemOpen
  \bibfield  {author} {\bibinfo {author} {\bibfnamefont {D.}~\bibnamefont
  {Nickelsen}}\ and\ \bibinfo {author} {\bibfnamefont {A.}~\bibnamefont
  {Engel}},\ }\href {\doibase 10.1140/epjb/e2011-20133-y} {\bibfield  {journal}
  {\bibinfo  {journal} {Eur. Phys. J. B}\ }\textbf {\bibinfo {volume} {82}},\
  \bibinfo {pages} {207–218} (\bibinfo {year} {2011})}\BibitemShut {NoStop}%
\bibitem [{\citenamefont {Nickelsen}\ and\ \citenamefont
  {Engel}(2012)}]{nickelsen2012}%
  \BibitemOpen
  \bibfield  {author} {\bibinfo {author} {\bibfnamefont {D.}~\bibnamefont
  {Nickelsen}}\ and\ \bibinfo {author} {\bibfnamefont {A.}~\bibnamefont
  {Engel}},\ }\href {\doibase 10.1088/0031-8949/86/05/058503} {\bibfield
  {journal} {\bibinfo  {journal} {Physica Scripta}\ }\textbf {\bibinfo {volume}
  {86}},\ \bibinfo {pages} {058503} (\bibinfo {year} {2012})}\BibitemShut
  {NoStop}%
\bibitem [{\citenamefont {Hoppenau}\ and\ \citenamefont
  {Engel}(2013)}]{hoppenau2013}%
  \BibitemOpen
  \bibfield  {author} {\bibinfo {author} {\bibfnamefont {J.}~\bibnamefont
  {Hoppenau}}\ and\ \bibinfo {author} {\bibfnamefont {A.}~\bibnamefont
  {Engel}},\ }\href {\doibase 10.1088/1742-5468/2013/06/P06004} {\bibfield
  {journal} {\bibinfo  {journal} {J. Stat. Mech.}\ }\textbf {\bibinfo {volume}
  {2013}},\ \bibinfo {pages} {P06004} (\bibinfo {year} {2013})}\BibitemShut
  {NoStop}%
\bibitem [{\citenamefont {Kwon}\ \emph {et~al.}(2013)\citenamefont {Kwon},
  \citenamefont {Noh},\ and\ \citenamefont {Park}}]{kwon2013}%
  \BibitemOpen
  \bibfield  {author} {\bibinfo {author} {\bibfnamefont {C.}~\bibnamefont
  {Kwon}}, \bibinfo {author} {\bibfnamefont {J.~D.}\ \bibnamefont {Noh}}, \
  and\ \bibinfo {author} {\bibfnamefont {H.}~\bibnamefont {Park}},\ }\href
  {\doibase 10.1103/PhysRevE.88.062102} {\bibfield  {journal} {\bibinfo
  {journal} {Phys. Rev. E}\ }\textbf {\bibinfo {volume} {88}},\ \bibinfo
  {pages} {062102} (\bibinfo {year} {2013})}\BibitemShut {NoStop}%
\bibitem [{\citenamefont {Ryabov}\ \emph {et~al.}(2013)\citenamefont {Ryabov},
  \citenamefont {Dierl}, \citenamefont {Chvosta}, \citenamefont {Einax},\ and\
  \citenamefont {Maass}}]{ryabov2013}%
  \BibitemOpen
  \bibfield  {author} {\bibinfo {author} {\bibfnamefont {A.}~\bibnamefont
  {Ryabov}}, \bibinfo {author} {\bibfnamefont {M.}~\bibnamefont {Dierl}},
  \bibinfo {author} {\bibfnamefont {P.}~\bibnamefont {Chvosta}}, \bibinfo
  {author} {\bibfnamefont {M.}~\bibnamefont {Einax}}, \ and\ \bibinfo {author}
  {\bibfnamefont {P.}~\bibnamefont {Maass}},\ }\href {\doibase
  10.1088/1751-8113/46/7/075002} {\bibfield  {journal} {\bibinfo  {journal} {J.
  Phys. A}\ }\textbf {\bibinfo {volume} {46}},\ \bibinfo {pages} {075002}
  (\bibinfo {year} {2013})}\BibitemShut {NoStop}%
\bibitem [{\citenamefont {Holubec}\ \emph {et~al.}(2015)\citenamefont
  {Holubec}, \citenamefont {Dierl}, \citenamefont {Einax}, \citenamefont
  {Maass}, \citenamefont {Chvosta},\ and\ \citenamefont
  {Ryabov}}]{holubec2015}%
  \BibitemOpen
  \bibfield  {author} {\bibinfo {author} {\bibfnamefont {V.}~\bibnamefont
  {Holubec}}, \bibinfo {author} {\bibfnamefont {M.}~\bibnamefont {Dierl}},
  \bibinfo {author} {\bibfnamefont {M.}~\bibnamefont {Einax}}, \bibinfo
  {author} {\bibfnamefont {P.}~\bibnamefont {Maass}}, \bibinfo {author}
  {\bibfnamefont {P.}~\bibnamefont {Chvosta}}, \ and\ \bibinfo {author}
  {\bibfnamefont {A.}~\bibnamefont {Ryabov}},\ }\href {\doibase
  10.1088/0031-8949/2015/T165/014024} {\bibfield  {journal} {\bibinfo
  {journal} {Physica Scripta}\ }\textbf {\bibinfo {volume} {2015}},\ \bibinfo
  {pages} {014024} (\bibinfo {year} {2015})}\BibitemShut {NoStop}%
\bibitem [{\citenamefont {Saha}\ and\ \citenamefont
  {Mukherji}(2015)}]{saha2015}%
  \BibitemOpen
  \bibfield  {author} {\bibinfo {author} {\bibfnamefont {B.}~\bibnamefont
  {Saha}}\ and\ \bibinfo {author} {\bibfnamefont {S.}~\bibnamefont
  {Mukherji}},\ }\href {\doibase 10.1140/epjb/e2015-60179-1} {\bibfield
  {journal} {\bibinfo  {journal} {Eur. Physi. J. B}\ }\textbf {\bibinfo
  {volume} {88}},\ \bibinfo {pages} {146} (\bibinfo {year} {2015})}\BibitemShut
  {NoStop}%
\bibitem [{\citenamefont {Chvosta}\ \emph {et~al.}(2020)\citenamefont
  {Chvosta}, \citenamefont {Lips}, \citenamefont {Holubec}, \citenamefont
  {Ryabov},\ and\ \citenamefont {Maass}}]{chvosta2020}%
  \BibitemOpen
  \bibfield  {author} {\bibinfo {author} {\bibfnamefont {P.}~\bibnamefont
  {Chvosta}}, \bibinfo {author} {\bibfnamefont {D.}~\bibnamefont {Lips}},
  \bibinfo {author} {\bibfnamefont {V.}~\bibnamefont {Holubec}}, \bibinfo
  {author} {\bibfnamefont {A.}~\bibnamefont {Ryabov}}, \ and\ \bibinfo {author}
  {\bibfnamefont {P.}~\bibnamefont {Maass}},\ }\href {\doibase
  10.1088/1751-8121/ab95c2} {\bibfield  {journal} {\bibinfo  {journal} {J.
  Phys. A}\ }\textbf {\bibinfo {volume} {53}},\ \bibinfo {pages} {275001}
  (\bibinfo {year} {2020})}\BibitemShut {NoStop}%
\bibitem [{\citenamefont {Verley}\ \emph {et~al.}(2014)\citenamefont {Verley},
  \citenamefont {den Broeck},\ and\ \citenamefont {Esposito}}]{verley2014}%
  \BibitemOpen
  \bibfield  {author} {\bibinfo {author} {\bibfnamefont {G.}~\bibnamefont
  {Verley}}, \bibinfo {author} {\bibfnamefont {C.~V.}\ \bibnamefont {den
  Broeck}}, \ and\ \bibinfo {author} {\bibfnamefont {M.}~\bibnamefont
  {Esposito}},\ }\href {\doibase 10.1088/1367-2630/16/9/095001} {\bibfield
  {journal} {\bibinfo  {journal} {New J. Phys.}\ }\textbf {\bibinfo {volume}
  {16}},\ \bibinfo {pages} {095001} (\bibinfo {year} {2014})}\BibitemShut
  {NoStop}%
\bibitem [{\citenamefont {Manikandan}\ and\ \citenamefont
  {Krishnamurthy}(2017)}]{manikandan2017}%
  \BibitemOpen
  \bibfield  {author} {\bibinfo {author} {\bibfnamefont {S.}~\bibnamefont
  {Manikandan}}\ and\ \bibinfo {author} {\bibfnamefont {S.}~\bibnamefont
  {Krishnamurthy}},\ }\href {\doibase 10.1140/epjb/e2017-80432-9} {\bibfield
  {journal} {\bibinfo  {journal} {Eur. Phys. J. B}\ }\textbf {\bibinfo {volume}
  {90}},\ \bibinfo {pages} {258} (\bibinfo {year} {2017})}\BibitemShut
  {NoStop}%
\bibitem [{\citenamefont {Funo}\ and\ \citenamefont {Quan}(2018)}]{funo2018}%
  \BibitemOpen
  \bibfield  {author} {\bibinfo {author} {\bibfnamefont {K.}~\bibnamefont
  {Funo}}\ and\ \bibinfo {author} {\bibfnamefont {H.~T.}\ \bibnamefont
  {Quan}},\ }\href {\doibase 10.1103/PhysRevLett.121.040602} {\bibfield
  {journal} {\bibinfo  {journal} {Phys. Rev. Lett.}\ }\textbf {\bibinfo
  {volume} {121}},\ \bibinfo {pages} {040602} (\bibinfo {year}
  {2018})}\BibitemShut {NoStop}%
\bibitem [{\citenamefont {Hoppenau}\ \emph {et~al.}(2013)\citenamefont
  {Hoppenau}, \citenamefont {Niemann},\ and\ \citenamefont
  {Engel}}]{hoppenau2013engine}%
  \BibitemOpen
  \bibfield  {author} {\bibinfo {author} {\bibfnamefont {J.}~\bibnamefont
  {Hoppenau}}, \bibinfo {author} {\bibfnamefont {M.}~\bibnamefont {Niemann}}, \
  and\ \bibinfo {author} {\bibfnamefont {A.}~\bibnamefont {Engel}},\ }\href
  {\doibase 10.1103/PhysRevE.87.062127} {\bibfield  {journal} {\bibinfo
  {journal} {Phys. Rev. E}\ }\textbf {\bibinfo {volume} {87}},\ \bibinfo
  {pages} {062127} (\bibinfo {year} {2013})}\BibitemShut {NoStop}%
\bibitem [{\citenamefont {Holubec}(2014)}]{holubec2014}%
  \BibitemOpen
  \bibfield  {author} {\bibinfo {author} {\bibfnamefont {V.}~\bibnamefont
  {Holubec}},\ }\href {\doibase 10.1088/1742-5468/2014/05/P05022} {\bibfield
  {journal} {\bibinfo  {journal} {J. Stat. Mech.}\ }\textbf {\bibinfo {volume}
  {2013}},\ \bibinfo {pages} {P05022} (\bibinfo {year} {2014})}\BibitemShut
  {NoStop}%
\bibitem [{\citenamefont {Proesmans}\ \emph {et~al.}(2015)\citenamefont
  {Proesmans}, \citenamefont {Driesen}, \citenamefont {Cleuren},\ and\
  \citenamefont {Van~den Broeck}}]{proesmans2015}%
  \BibitemOpen
  \bibfield  {author} {\bibinfo {author} {\bibfnamefont {K.}~\bibnamefont
  {Proesmans}}, \bibinfo {author} {\bibfnamefont {C.}~\bibnamefont {Driesen}},
  \bibinfo {author} {\bibfnamefont {B.}~\bibnamefont {Cleuren}}, \ and\
  \bibinfo {author} {\bibfnamefont {C.}~\bibnamefont {Van~den Broeck}},\ }\href
  {\doibase 10.1103/PhysRevE.92.032105} {\bibfield  {journal} {\bibinfo
  {journal} {Phys. Rev. E}\ }\textbf {\bibinfo {volume} {92}},\ \bibinfo
  {pages} {032105} (\bibinfo {year} {2015})}\BibitemShut {NoStop}%
\bibitem [{\citenamefont {Holubec}\ and\ \citenamefont
  {Ryabov}(2021)}]{holubec2022}%
  \BibitemOpen
  \bibfield  {author} {\bibinfo {author} {\bibfnamefont {V.}~\bibnamefont
  {Holubec}}\ and\ \bibinfo {author} {\bibfnamefont {A.}~\bibnamefont
  {Ryabov}},\ }\href {\doibase 10.1088/1751-8121/ac3aac} {\bibfield  {journal}
  {\bibinfo  {journal} {J. Phys. A: Math. Theor.}\ }\textbf {\bibinfo {volume}
  {55}},\ \bibinfo {pages} {013001} (\bibinfo {year} {2021})}\BibitemShut
  {NoStop}%
\bibitem [{\citenamefont {Vucelja}\ \emph {et~al.}(2015)\citenamefont
  {Vucelja}, \citenamefont {Turitsyn},\ and\ \citenamefont
  {Chertkov}}]{vucelja2015}%
  \BibitemOpen
  \bibfield  {author} {\bibinfo {author} {\bibfnamefont {M.}~\bibnamefont
  {Vucelja}}, \bibinfo {author} {\bibfnamefont {K.~S.}\ \bibnamefont
  {Turitsyn}}, \ and\ \bibinfo {author} {\bibfnamefont {M.}~\bibnamefont
  {Chertkov}},\ }\href {\doibase 10.1103/PhysRevE.91.022123} {\bibfield
  {journal} {\bibinfo  {journal} {Phys. Rev. E}\ }\textbf {\bibinfo {volume}
  {91}},\ \bibinfo {pages} {022123} (\bibinfo {year} {2015})}\BibitemShut
  {NoStop}%
\bibitem [{\citenamefont {Damak}\ \emph {et~al.}(2020)\citenamefont {Damak},
  \citenamefont {Hammami},\ and\ \citenamefont {Pillet}}]{damak2020}%
  \BibitemOpen
  \bibfield  {author} {\bibinfo {author} {\bibfnamefont {M.}~\bibnamefont
  {Damak}}, \bibinfo {author} {\bibfnamefont {M.}~\bibnamefont {Hammami}}, \
  and\ \bibinfo {author} {\bibfnamefont {C.-A.}\ \bibnamefont {Pillet}},\
  }\href {\doibase 10.1007/s10955-019-02398-x} {\bibfield  {journal} {\bibinfo
  {journal} {J. Stat. Phys.}\ }\textbf {\bibinfo {volume} {180}},\ \bibinfo
  {pages} {263} (\bibinfo {year} {2020})}\BibitemShut {NoStop}%
\bibitem [{\citenamefont {Wu}\ \emph {et~al.}(2023)\citenamefont {Wu},
  \citenamefont {Chen}, \citenamefont {Pei}, \citenamefont {Zhang},\ and\
  \citenamefont {Quan}}]{wu2023}%
  \BibitemOpen
  \bibfield  {author} {\bibinfo {author} {\bibfnamefont {Y.-X.}\ \bibnamefont
  {Wu}}, \bibinfo {author} {\bibfnamefont {J.-F.}\ \bibnamefont {Chen}},
  \bibinfo {author} {\bibfnamefont {J.-H.}\ \bibnamefont {Pei}}, \bibinfo
  {author} {\bibfnamefont {F.}~\bibnamefont {Zhang}}, \ and\ \bibinfo {author}
  {\bibfnamefont {H.~T.}\ \bibnamefont {Quan}},\ }\href {\doibase
  10.1103/PhysRevE.107.064115} {\bibfield  {journal} {\bibinfo  {journal}
  {Phys. Rev. E}\ }\textbf {\bibinfo {volume} {107}},\ \bibinfo {pages}
  {064115} (\bibinfo {year} {2023})}\BibitemShut {NoStop}%
\bibitem [{\citenamefont {Gupta}\ and\ \citenamefont
  {Sivak}(2021)}]{gupta2021}%
  \BibitemOpen
  \bibfield  {author} {\bibinfo {author} {\bibfnamefont {D.}~\bibnamefont
  {Gupta}}\ and\ \bibinfo {author} {\bibfnamefont {D.~A.}\ \bibnamefont
  {Sivak}},\ }\href {\doibase 10.1103/PhysRevE.104.024605} {\bibfield
  {journal} {\bibinfo  {journal} {Phys. Rev. E}\ }\textbf {\bibinfo {volume}
  {104}},\ \bibinfo {pages} {024605} (\bibinfo {year} {2021})}\BibitemShut
  {NoStop}%
\bibitem [{\citenamefont {Gibbs}\ and\ \citenamefont
  {DiMarzio}(1959)}]{gibbs1959}%
  \BibitemOpen
  \bibfield  {author} {\bibinfo {author} {\bibfnamefont {J.~H.}\ \bibnamefont
  {Gibbs}}\ and\ \bibinfo {author} {\bibfnamefont {E.~A.}\ \bibnamefont
  {DiMarzio}},\ }\href {\doibase 10.1063/1.1729886} {\bibfield  {journal}
  {\bibinfo  {journal} {J. Chem. Phys.}\ }\textbf {\bibinfo {volume} {30}},\
  \bibinfo {pages} {271} (\bibinfo {year} {1959})}\BibitemShut {NoStop}%
\bibitem [{\citenamefont {Applequist}\ and\ \citenamefont
  {Damle}(1953)}]{applequist1963}%
  \BibitemOpen
  \bibfield  {author} {\bibinfo {author} {\bibfnamefont {J.}~\bibnamefont
  {Applequist}}\ and\ \bibinfo {author} {\bibfnamefont {V.}~\bibnamefont
  {Damle}},\ }\href {\doibase 10.1063/1.1734089} {\bibfield  {journal}
  {\bibinfo  {journal} {J. Chem. Phys.}\ }\textbf {\bibinfo {volume} {39}},\
  \bibinfo {pages} {2719} (\bibinfo {year} {1953})}\BibitemShut {NoStop}%
\bibitem [{\citenamefont {Bockelmann}\ \emph {et~al.}(1997)\citenamefont
  {Bockelmann}, \citenamefont {Essevaz-Roulet},\ and\ \citenamefont
  {Heslot}}]{bockelmann1997}%
  \BibitemOpen
  \bibfield  {author} {\bibinfo {author} {\bibfnamefont {U.}~\bibnamefont
  {Bockelmann}}, \bibinfo {author} {\bibfnamefont {B.}~\bibnamefont
  {Essevaz-Roulet}}, \ and\ \bibinfo {author} {\bibfnamefont {F.}~\bibnamefont
  {Heslot}},\ }\href {\doibase 10.1103/PhysRevLett.79.4489} {\bibfield
  {journal} {\bibinfo  {journal} {Phys. Rev. Lett.}\ }\textbf {\bibinfo
  {volume} {79}},\ \bibinfo {pages} {4489} (\bibinfo {year}
  {1997})}\BibitemShut {NoStop}%
\bibitem [{\citenamefont {Marenduzzo}\ \emph {et~al.}(2001)\citenamefont
  {Marenduzzo}, \citenamefont {Trovato},\ and\ \citenamefont
  {Maritan}}]{marenduzzo2001}%
  \BibitemOpen
  \bibfield  {author} {\bibinfo {author} {\bibfnamefont {D.}~\bibnamefont
  {Marenduzzo}}, \bibinfo {author} {\bibfnamefont {A.}~\bibnamefont {Trovato}},
  \ and\ \bibinfo {author} {\bibfnamefont {A.}~\bibnamefont {Maritan}},\ }\href
  {\doibase 10.1103/PhysRevE.64.031901} {\bibfield  {journal} {\bibinfo
  {journal} {Phys. Rev. E}\ }\textbf {\bibinfo {volume} {64}},\ \bibinfo
  {pages} {031901} (\bibinfo {year} {2001})}\BibitemShut {NoStop}%
\bibitem [{\citenamefont {Lubensky}\ and\ \citenamefont
  {Nelson}(2000)}]{lubensky2000}%
  \BibitemOpen
  \bibfield  {author} {\bibinfo {author} {\bibfnamefont {D.~K.}\ \bibnamefont
  {Lubensky}}\ and\ \bibinfo {author} {\bibfnamefont {D.~R.}\ \bibnamefont
  {Nelson}},\ }\href {\doibase 10.1103/PhysRevLett.85.1572} {\bibfield
  {journal} {\bibinfo  {journal} {Phys. Rev. Lett.}\ }\textbf {\bibinfo
  {volume} {85}},\ \bibinfo {pages} {1572} (\bibinfo {year}
  {2000})}\BibitemShut {NoStop}%
\bibitem [{\citenamefont {Hwa}\ \emph {et~al.}(2003)\citenamefont {Hwa},
  \citenamefont {Marinari}, \citenamefont {Sneppen},\ and\ \citenamefont
  {Tang}}]{hwa2003}%
  \BibitemOpen
  \bibfield  {author} {\bibinfo {author} {\bibfnamefont {T.}~\bibnamefont
  {Hwa}}, \bibinfo {author} {\bibfnamefont {E.}~\bibnamefont {Marinari}},
  \bibinfo {author} {\bibfnamefont {K.}~\bibnamefont {Sneppen}}, \ and\
  \bibinfo {author} {\bibfnamefont {L.}~\bibnamefont {Tang}},\ }\href@noop {}
  {\bibfield  {journal} {\bibinfo  {journal} {PNAS}\ }\textbf {\bibinfo
  {volume} {100}},\ \bibinfo {pages} {4411} (\bibinfo {year}
  {2003})}\BibitemShut {NoStop}%
\bibitem [{\citenamefont {Kafri}\ \emph {et~al.}(2000)\citenamefont {Kafri},
  \citenamefont {Mukamel},\ and\ \citenamefont {Peliti}}]{Kafri2000}%
  \BibitemOpen
  \bibfield  {author} {\bibinfo {author} {\bibfnamefont {Y.}~\bibnamefont
  {Kafri}}, \bibinfo {author} {\bibfnamefont {D.}~\bibnamefont {Mukamel}}, \
  and\ \bibinfo {author} {\bibfnamefont {L.}~\bibnamefont {Peliti}},\ }\href
  {\doibase 10.1103/PhysRevLett.85.4988} {\bibfield  {journal} {\bibinfo
  {journal} {Phys. Rev. Lett.}\ }\textbf {\bibinfo {volume} {85}},\ \bibinfo
  {pages} {4988} (\bibinfo {year} {2000})}\BibitemShut {NoStop}%
\bibitem [{\citenamefont {Kafri}\ \emph
  {et~al.}(2002{\natexlab{a}})\citenamefont {Kafri}, \citenamefont {Mukamel},\
  and\ \citenamefont {Peliti}}]{Kafri2002}%
  \BibitemOpen
  \bibfield  {author} {\bibinfo {author} {\bibfnamefont {Y.}~\bibnamefont
  {Kafri}}, \bibinfo {author} {\bibfnamefont {D.}~\bibnamefont {Mukamel}}, \
  and\ \bibinfo {author} {\bibfnamefont {L.}~\bibnamefont {Peliti}},\ }\href
  {\doibase 10.1140/epjb/e20020138} {\bibfield  {journal} {\bibinfo  {journal}
  {Eur. Phys. J. B}\ }\textbf {\bibinfo {volume} {27}},\ \bibinfo {pages}
  {135–146} (\bibinfo {year} {2002}{\natexlab{a}})}\BibitemShut {NoStop}%
\bibitem [{\citenamefont {Kafri}\ \emph
  {et~al.}(2002{\natexlab{b}})\citenamefont {Kafri}, \citenamefont {Mukamel},\
  and\ \citenamefont {Peliti}}]{Kafri2002b}%
  \BibitemOpen
  \bibfield  {author} {\bibinfo {author} {\bibfnamefont {Y.}~\bibnamefont
  {Kafri}}, \bibinfo {author} {\bibfnamefont {D.}~\bibnamefont {Mukamel}}, \
  and\ \bibinfo {author} {\bibfnamefont {L.}~\bibnamefont {Peliti}},\ }\href
  {\doibase 10.1016/S0378-4371(02)00483-1} {\bibfield  {journal} {\bibinfo
  {journal} {Physica A}\ }\textbf {\bibinfo {volume} {306}},\ \bibinfo {pages}
  {39} (\bibinfo {year} {2002}{\natexlab{b}})}\BibitemShut {NoStop}%
\bibitem [{\citenamefont {Deger}\ \emph {et~al.}(2018)\citenamefont {Deger},
  \citenamefont {Brandner},\ and\ \citenamefont {Flindt}}]{Deger2018}%
  \BibitemOpen
  \bibfield  {author} {\bibinfo {author} {\bibfnamefont {A.}~\bibnamefont
  {Deger}}, \bibinfo {author} {\bibfnamefont {K.}~\bibnamefont {Brandner}}, \
  and\ \bibinfo {author} {\bibfnamefont {C.}~\bibnamefont {Flindt}},\ }\href
  {\doibase 10.1103/PhysRevE.97.012115} {\bibfield  {journal} {\bibinfo
  {journal} {Phys. Rev. E}\ }\textbf {\bibinfo {volume} {97}},\ \bibinfo
  {pages} {012115} (\bibinfo {year} {2018})}\BibitemShut {NoStop}%
\bibitem [{\citenamefont {Roland}\ \emph {et~al.}(2009)\citenamefont {Roland},
  \citenamefont {Hatch}, \citenamefont {Prentiss},\ and\ \citenamefont
  {Shakhnovich}}]{roland2009}%
  \BibitemOpen
  \bibfield  {author} {\bibinfo {author} {\bibfnamefont {C.~B.}\ \bibnamefont
  {Roland}}, \bibinfo {author} {\bibfnamefont {K.~A.}\ \bibnamefont {Hatch}},
  \bibinfo {author} {\bibfnamefont {M.}~\bibnamefont {Prentiss}}, \ and\
  \bibinfo {author} {\bibfnamefont {E.~I.}\ \bibnamefont {Shakhnovich}},\
  }\href {\doibase 10.1103/PhysRevE.79.051923} {\bibfield  {journal} {\bibinfo
  {journal} {Phys. Rev. E}\ }\textbf {\bibinfo {volume} {79}},\ \bibinfo
  {pages} {051923} (\bibinfo {year} {2009})}\BibitemShut {NoStop}%
\bibitem [{\citenamefont {Kafri}\ and\ \citenamefont
  {Polkovnikov}(2006)}]{Kafri2006}%
  \BibitemOpen
  \bibfield  {author} {\bibinfo {author} {\bibfnamefont {Y.}~\bibnamefont
  {Kafri}}\ and\ \bibinfo {author} {\bibfnamefont {A.}~\bibnamefont
  {Polkovnikov}},\ }\href {\doibase 10.1103/PhysRevLett.97.208104} {\bibfield
  {journal} {\bibinfo  {journal} {Phys. Rev. Lett.}\ }\textbf {\bibinfo
  {volume} {97}},\ \bibinfo {pages} {208104} (\bibinfo {year}
  {2006})}\BibitemShut {NoStop}%
\bibitem [{\citenamefont {Sauer-Budge}\ \emph {et~al.}(2003)\citenamefont
  {Sauer-Budge}, \citenamefont {Nyamwanda}, \citenamefont {Lubensky},\ and\
  \citenamefont {Branton}}]{sauer-budge2003}%
  \BibitemOpen
  \bibfield  {author} {\bibinfo {author} {\bibfnamefont {A.~F.}\ \bibnamefont
  {Sauer-Budge}}, \bibinfo {author} {\bibfnamefont {J.~A.}\ \bibnamefont
  {Nyamwanda}}, \bibinfo {author} {\bibfnamefont {D.~K.}\ \bibnamefont
  {Lubensky}}, \ and\ \bibinfo {author} {\bibfnamefont {D.}~\bibnamefont
  {Branton}},\ }\href {\doibase 10.1103/PhysRevLett.90.238101} {\bibfield
  {journal} {\bibinfo  {journal} {Phys. Rev. Lett.}\ }\textbf {\bibinfo
  {volume} {90}},\ \bibinfo {pages} {238101} (\bibinfo {year}
  {2003})}\BibitemShut {NoStop}%
\bibitem [{\citenamefont {Cocco}\ \emph {et~al.}(2002)\citenamefont {Cocco},
  \citenamefont {Marko},\ and\ \citenamefont {Monasson}}]{cocco2002}%
  \BibitemOpen
  \bibfield  {author} {\bibinfo {author} {\bibfnamefont {S.}~\bibnamefont
  {Cocco}}, \bibinfo {author} {\bibfnamefont {J.~F.}\ \bibnamefont {Marko}}, \
  and\ \bibinfo {author} {\bibfnamefont {R.}~\bibnamefont {Monasson}},\ }\href
  {\doibase https://doi.org/10.1016/S1631-0705(02)01345-2} {\bibfield
  {journal} {\bibinfo  {journal} {Comp. Rend. Phys.}\ }\textbf {\bibinfo
  {volume} {3}},\ \bibinfo {pages} {569} (\bibinfo {year} {2002})}\BibitemShut
  {NoStop}%
\bibitem [{\citenamefont {Rissone}\ and\ \citenamefont
  {Ritort}(2022)}]{rissone2022}%
  \BibitemOpen
  \bibfield  {author} {\bibinfo {author} {\bibfnamefont {P.}~\bibnamefont
  {Rissone}}\ and\ \bibinfo {author} {\bibfnamefont {F.}~\bibnamefont
  {Ritort}},\ }\href {\doibase 10.3390/life12071089} {\bibfield  {journal}
  {\bibinfo  {journal} {Life}\ }\textbf {\bibinfo {volume} {12}},\ \bibinfo
  {pages} {1089} (\bibinfo {year} {2022})}\BibitemShut {NoStop}%
\bibitem [{\citenamefont {M\"{u}ller}\ \emph {et~al.}(2002)\citenamefont
  {M\"{u}ller}, \citenamefont {Krzakala},\ and\ \citenamefont
  {M\'ezard}}]{mueller2002}%
  \BibitemOpen
  \bibfield  {author} {\bibinfo {author} {\bibfnamefont {M.}~\bibnamefont
  {M\"{u}ller}}, \bibinfo {author} {\bibfnamefont {F.}~\bibnamefont
  {Krzakala}}, \ and\ \bibinfo {author} {\bibfnamefont {M.}~\bibnamefont
  {M\'ezard}},\ }\href {\doibase 10.1140/epje/i2002-10057-5} {\bibfield
  {journal} {\bibinfo  {journal} {Eur. Phys. J. E}\ }\textbf {\bibinfo {volume}
  {9}},\ \bibinfo {pages} {67} (\bibinfo {year} {2002})}\BibitemShut {NoStop}%
\bibitem [{\citenamefont {Hartmann}(2015)}]{practical_guide2015}%
  \BibitemOpen
  \bibfield  {author} {\bibinfo {author} {\bibfnamefont {A.~K.}\ \bibnamefont
  {Hartmann}},\ }\href@noop {} {\emph {\bibinfo {title} {{Big Practical Guide
  to Computer Simulations}}}}\ (\bibinfo  {publisher} {World Scientific},\
  \bibinfo {address} {Singapore},\ \bibinfo {year} {2015})\BibitemShut
  {NoStop}%
\bibitem [{\citenamefont {Hartmann}(2011)}]{largest-2011}%
  \BibitemOpen
  \bibfield  {author} {\bibinfo {author} {\bibfnamefont {A.~K.}\ \bibnamefont
  {Hartmann}},\ }\href@noop {} {\bibfield  {journal} {\bibinfo  {journal} {Eur.
  Phys. J. B}\ }\textbf {\bibinfo {volume} {84}},\ \bibinfo {pages} {627}
  (\bibinfo {year} {2011})}\BibitemShut {NoStop}%
\bibitem [{\citenamefont {Ferrenberg}\ and\ \citenamefont
  {Swendsen}(1989)}]{Ferrenberg_1989}%
  \BibitemOpen
  \bibfield  {author} {\bibinfo {author} {\bibfnamefont {A.~M.}\ \bibnamefont
  {Ferrenberg}}\ and\ \bibinfo {author} {\bibfnamefont {R.~H.}\ \bibnamefont
  {Swendsen}},\ }\href {\doibase 10.1103/PhysRevLett.63.1195} {\bibfield
  {journal} {\bibinfo  {journal} {Phys. Rev. Lett.}\ }\textbf {\bibinfo
  {volume} {63}},\ \bibinfo {pages} {1195} (\bibinfo {year}
  {1989})}\BibitemShut {NoStop}%
\bibitem [{\citenamefont {Werner}()}]{werner2022}%
  \BibitemOpen
  \bibfield  {author} {\bibinfo {author} {\bibfnamefont {P.}~\bibnamefont
  {Werner}},\ }\href@noop {} {\enquote {\bibinfo {title} {A software tool for
  "gluing" distributions},}\ }\Eprint {http://arxiv.org/abs/2207.08429}
  {arXiv:2207.08429 [physics.data-an]} \BibitemShut {NoStop}%
\end{thebibliography}%

\end{document}